\documentclass[10pt, twocolumn]{IEEEtran}
\usepackage{indentfirst}
\usepackage{cite}
\usepackage{amsmath}
\usepackage{amsfonts}
\usepackage{amssymb}
\usepackage{times}
\usepackage{subfigure}
\usepackage[english]{babel}
\usepackage{stfloats}
\usepackage{dsfont}
\usepackage[dvips]{graphicx}
\usepackage{array}
\usepackage[normalem]{ulem}
\usepackage{epsf}
\usepackage{psfrag}
\usepackage{comment}

\newcommand{\figw}{0.98\columnwidth}

\newcommand{\bi}{\begin{itemize}}
\newcommand{\ei}{\end{itemize}}
\newcommand{\ben}{\begin{enumerate}}
\newcommand{\een}{\end{enumerate}}
\newcommand{\bc}{\begin{cases}}
\newcommand{\ec}{\end{cases}}
\newcommand{\bd}{\begin{description}}
\newcommand{\ed}{\end{description}}

\newcommand{\be}{\begin{equation}}
\newcommand{\ee}{\end{equation}}
\newcommand{\bea}{\begin{eqnarray}}
\newcommand{\eea}{\end{eqnarray}}

\newcommand{\vup}{\vspace{-1mm}}
\newtheorem{theorem}{Theorem}
\newtheorem{lemma}{Lemma}

\newcounter{mytempeqncnt}

\DeclareFontFamily{U}{futm}{}
\DeclareFontShape{U}{futm}{m}{n}{
 <-> s * [0.98] fourier-bb
 }{}
\DeclareSymbolFont{leoBB}{U}{futm}{m}{n}
\DeclareSymbolFontAlphabet{\mathreal}{leoBB}
\AtBeginDocument

\begin{document}

\title{Cognitive Interference Management in Retransmission-Based Wireless Networks}
\author{ \large Marco Levorato$^{*}$, \emph{Member, IEEE}, Urbashi Mitra$^\dag$, \emph{Fellow, IEEE}, Michele Zorzi$^{\ddag}$, \emph{Fellow, IEEE}
\thanks{Prof. Mitra's and Marco Levorato's work was supported in part by the National Science Foundation (NSF) under Grant CCF-0917343. Prof. Zorzi's work was partially supported by the European Commission under the SAPHYRE project (FP7 ICT-248001), and by the US Army Research Office through Grant No. W911NF-09-1-0456.
Part of this work was presented at the Forty-Seventh Annual Allerton Conference on Communication, Control, and Computing, Monticello, IL, Sep. 30 - Oct 2, 2009. 
Marco Levorato is with the Dept. of Electrical Engineering, Stanford University, Stanford, CA 94305 USA and the Dept. of Electrical Engineering,
University of Southern California, Los Angeles, CA  90089 USA (e-mail: levorato@stanford.edu). Urbashi Mitra is with the Dept. of Electrical Engineering,
University of Southern California, Los Angeles, CA  90089 USA (e-mail: ubli@usc.edu). Michele Zorzi is with the Dept.\ of Information Engineering, University of Padova, 35131 Padova, Italy, and the California Institute of Telecommunications and Information Technology, University of California at San Diego, La Jolla, CA 92093 USA(e-mail: zorzi@dei.unipd.it).}
}
\date{}

\maketitle

\pagestyle{empty}
\thispagestyle{empty}

\begin{abstract}
Cognitive radio methodologies have the potential to dramatically increase the throughput of wireless systems.
Herein, control strategies which enable the superposition in time and frequency of primary and secondary user
transmissions are explored in contrast to more traditional sensing approaches which only allow the secondary
user to transmit when the primary user is idle. In this work, the
optimal transmission policy for the secondary user when the primary user adopts
a retransmission based error control scheme is investigated. The policy aims to
maximize the secondary users' throughput, with a constraint on the
throughput loss and failure probability of the primary user. Due to the constraint,
the optimal policy is randomized, and determines how often the secondary user
transmits according to the retransmission state of the packet being served by the primary user.
The resulting optimal strategy of the secondary user is proven to have
a unique structure. In particular, the optimal throughput is achieved by the
secondary user by concentrating its transmission, and thus its interference to the primary user,
in the first transmissions of a primary user packet. The rather simple framework considered in this paper
highlights two fundamental aspects of cognitive networks that have not been covered so far: (i)
the networking mechanisms implemented by the primary users (error control by means of retransmissions
in the considered model) react to secondary users' activity; (ii) if networking mechanisms are considered,
then their \emph{state} must be taken into account when optimizing secondary users' strategy, \emph{i.e.},
a strategy based on a binary active/idle perception of the primary users' state is suboptimal.
\end{abstract}

\section{Introduction}
\label{intro}
Cognitive radio has been the subject of intense research of late, \emph{e.g.},~\cite{pap5,pap6,commag1,infoneely,wenyi,pap2,pap3,mc,mitola}, due to its potential to increase the efficiency of wireless networks. Unlicensed secondary users adapt their operations around those of the primary users and the surrounding network environment to opportunistically exploit available resources while limiting their interference with licensed primary users.
Most prior work~\cite{pap5,pap6,commag1,infoneely} focuses on a \emph{white space} approach, where the secondary users
sense the channel in order to detect time/frequency slots left unused by the primary users and exploit them for transmission.
Pure \emph{white space} approaches are based on a \emph{zero-interference} rationale, \emph{i.e.}, the objective of the secondary
user is to not interfere at all with the primary user.
However, sensing errors may lead to unwanted collisions, thus degrading the
throughput achieved by the latter. Typically, primary users are modeled via a fixed Markov chain tracking the idle-busy
channel state, irrespective of the operations of the secondary users, according to the general assumption that primary users
are \emph{dumb} and non-adaptive devices.
 
However, such a model may not always be accurate. For instance, a collision may force
a primary user to schedule a retransmission and enter a backoff period. As a consequence, a collision may modify the arrival
rate of the packets at the primary destination, while changing the characterization of the generated traffic (burstiness of idle/busy slots).
As a consequence, the interaction between the primary users and the secondary users must be considered when analyzing the network.  Additionally, the use of signal processing methods (multiuser detection and multiple-input multiple-output systems) enables the superposition of secondary transmissions over a primary transmission while achieving accurate decoding of the primary user packet.
 
There exists some prior literature investigating the coexistence in the same time/frequency band of primary and secondary users with a focus on physical layer methods for static scenarios~\cite{wenyi,pap2,pap3,mc,int1,intcanc}. A thorough discussion of spectrum sharing under performance constraints
from an information theoretic perspective can be found in~\cite{capgast}.
Those approaches, though valuable
in some broadcasting network scenarios, do not characterize the dynamic interaction between the two classes of users. In contrast, our prior work, which inspires the current paper, studies concurrent transmission by secondary and primary users in a highly dynamic environment \cite{infomio}.
We explicitly consider an interference mitigation scenario, where the secondary user is allowed to transmit concurrently to the primary user, with a constraint on the performance loss suffered by the latter, in terms of either a reduced throughput or an increased failure probability.

In this work, we study access control policies for secondary users in wireless networks where nodes implement
a retransmission-based error control scheme. Some prior literature has investigated networks of primary users 
implementing Automatic Retransmission reQuest (ARQ).
In~\cite{aria1}, the secondary user exploits the retransmissions of primary
user packets in order to achieve a higher transmission rate. In fact, the secondary receiver can potentially decode the
primary user's packet in the first transmission and then opportunistically cancel interference~\cite{intcanc} in the following retransmissions.
However, the framework in~\cite{aria1} does not consider the dynamics of the network and the bias in the channel availability generated by
interference. In~\cite{arq_bit}, Eswaran \emph{et al} propose a framework where the secondary user exploits ARQ
feedback to estimate the throughput loss of the primary user and tune the transmission policy accordingly, by using
information theoretic results. In~\cite{activelearning}, Zhang proposed a learning algorithm for a scenario in which the
primary user adapts the transmitted power in response to interference.

The contribution of the present paper is to introduce the \emph{reactive primary user} scenario, where the activity of the secondary users
biases the temporal evolution of the stochastic process tracking the state of the primary users. A Markov model is proposed and the optimization
problem is formulated as a constrained Markov Decision Process (MDP). The structure of the optimal policy is derived analytically for 
a specific case. We focus on a network with two mutually interfering links, one primary and one
secondary. In the framework considered, a packet may be retransmitted a finite number of times, due to
transmission failure, before being discarded by the transmitter. Packet arrivals at the primary user are modeled with a fixed probability
that an empty slot, \emph{i.e.}, a slot in which a retransmission is not scheduled, is accessed for transmission. 

We study the interference that the
secondary user causes to the primary user and how this interference impacts the retransmission process of the latter. We explicitly consider
an interference mitigation scenario, where the secondary user is allowed to transmit concurrently to the primary user, with a constraint
on the performance loss suffered by the latter, in terms of either a reduced throughput or an increased failure probability. Our analysis
is based on a detailed Markov model of the network, accounting for the distortion of the retransmission process caused by secondary user
transmissions. Remarkably, this simple model captures a fundamental aspect of cognitive networks: the control mechanisms 
implemented by the primary users \emph{react} to the activity of the secondary users. An accurate stochastic model for the activity of
the primary users should include this effect. In the model considered herein, interference
from the secondary source increases the probability that primary source's transmissions fail. As a consequence, retransmissions
are triggered more often, and the stochastic characterization of primary source's channel occupation changes. 
Moreover, as the motion law of the state of the primary user depends on both the action of the secondary user and
the current state of the primary user (the retransmission index of the packet being served in the considered model),
then the secondary users' strategy should be based on the state of the primary user. This means that a binary
active/idle representation of the state of the primary users leads to suboptimal policies.

In this framework, interference due to the activity of the secondary user not only reduces the instantaneous average revenue
collected by the primary user in each state of the Markov chain modeling the network, but also changes the transition probabilities,
and thus the steady-state distribution, of the chain. 

The optimization problem can be formalized through a Linear Program. Due to the constraint on the maximum performance loss
of the primary user, the solution is a randomized policy,
\emph{i.e.}, the optimal policy assigns a probability to each action in the action set given the state of
the underlying Markov process. As we focus on a binary transmission/idleness action set, the randomized policy simply determines
how often the secondary user transmits given the state of the network. 

This problem, though conceptually simple, unveils important
issues and general behaviors. As the primary user implements a retransmission-based error control mechanism, the activity of the
secondary user biases the retransmission process via interference. Interference at the primary receiver increases the failure probability of
primary user's transmissions. Therefore, due to the activity of the secondary user the average number of transmissions of a primary
user's packet gets larger, together with the average time required to return to primary user's idle state.
Interestingly, the increase of the average number of transmissions of primary user's packets depends on the index
of the interfered transmission. For instance, while interference from the secondary user in the first transmission
of the primary user's packets potentially leads to a significant increase of the number of transmissions per packet, transmission by
the secondary user in the last allowed transmission of primary user's packets does not increase the average number of transmissions
at all. Thus, as observed before, the impact of
secondary users' transmission in the various states critically depends on the state of the primary network. On the other hand, first transmissions
occur more frequently than last transmissions, and, thus, the overall throughput collected by the secondary user as a function of the strategy
greatly depends on the states in which it concentrates its transmissions. The interplay between the cost of the primary user and the reward
of the secondary one due to the modifications of the steady-state distribution determines the optimal strategy.

An important observation concerns the availability of time slots in which the primary user is idle, \emph{i.e.}, the white spaces.
As the primary user implements a retransmission-based error control mechanism, failed decoding at the primary receiver triggers
a further transmission of the packet, until the maximum number of transmissions per packet is reached. Therefore, the interference generated
by the activity of the secondary user increases the fraction of channel resource occupied by the primary user. This means that the
availability of white spaces decreases as the activity of the secondary user increases. This is an additional reason for carefully designing
the strategy of the secondary user.

If transmission by the primary user does not affect reception at the secondary receiver, and either
throughput or packet failure probability is considered as the metric for the primary user performance,
the optimal transmission strategy of the secondary user is shown to have a unique
structure. The throughput-optimal strategy concentrates transmissions by the secondary user in the region
of the state space corresponding to the first transmissions of a primary user's packet. According
to such optimal policy, the secondary user transmits with probability $1$ up to the $N_1$-th transmission of a
primary user's packet, with probability in $[0,1]$ in time slots in which the primary user is performing
the $N_1$-th transmission of a packet and with probability $0$ otherwise. The boundary state and the associated
transmission probability are determined to result into a bounded reduction of the time-average performance
of the primary user. This result also provides a simple algorithm to solve the linear program resulting
from the constrained optimization problem.


We also observe that the maximum \emph{aggressiveness} of the secondary user depends on the arrival rate at the primary user. In fact,
when the primary source spends most of its time idle, a longer retransmission process has a less deleterious effect on throughput.

The rest of the paper is organized as follows. Section~\ref{netdes} describes the network scenario considered throughout the paper.
Section~\ref{opt} defines the optimization problem and derives the Markov model of the network. In Section~\ref{optz},
the structure of the optimal strategy for the case in which primary users' transmission does not affect packet reception at the 
secondary receiver is derived. Section~\ref{gencase} discusses the optimal transmission policy for the
general case. In Section~\ref{numres}, numerical results highlighting the fundamental issues and behaviors described in the previous
sections are shown. Section~\ref{concl} concludes the paper.

\section{Network Description}
\label{netdes}
\begin{figure}[t]
	\centering
	\includegraphics[width=.5\columnwidth]{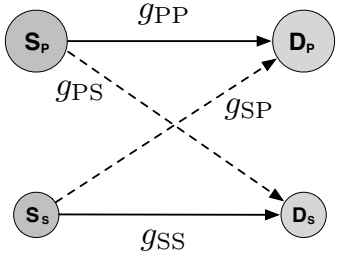}
\caption{Considered network. Direct links and interfering links are represented by solid and dashed arrows, respectively.}
\vup
\label{netfig}
\end{figure}
\begin{figure*}[!t]
	\centering
	\subfigure[\label{harq_sch}]{\includegraphics[width=.98\columnwidth]{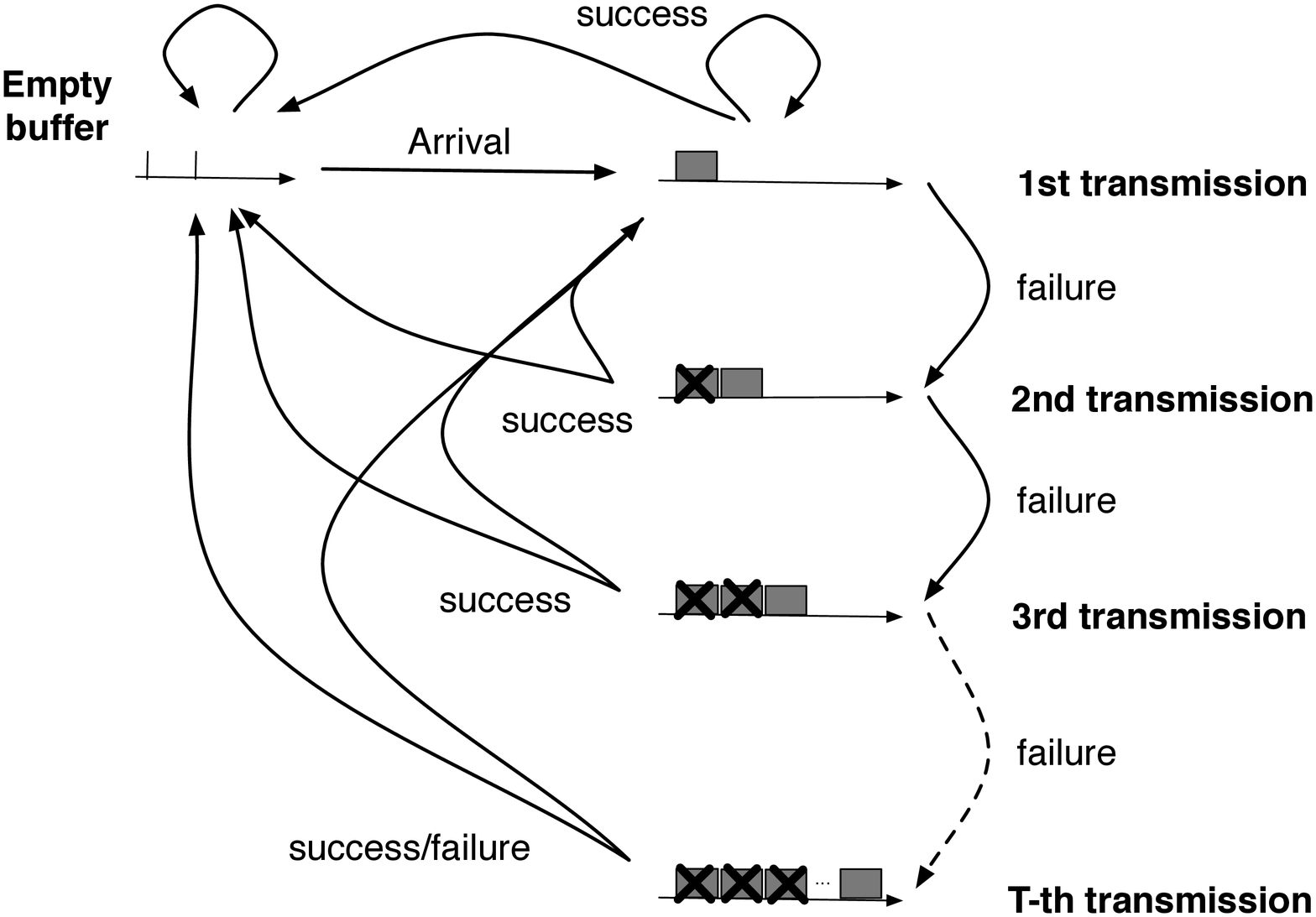}}
	\hfill
	\subfigure[\label{cat}]{\includegraphics[width=.98\columnwidth]{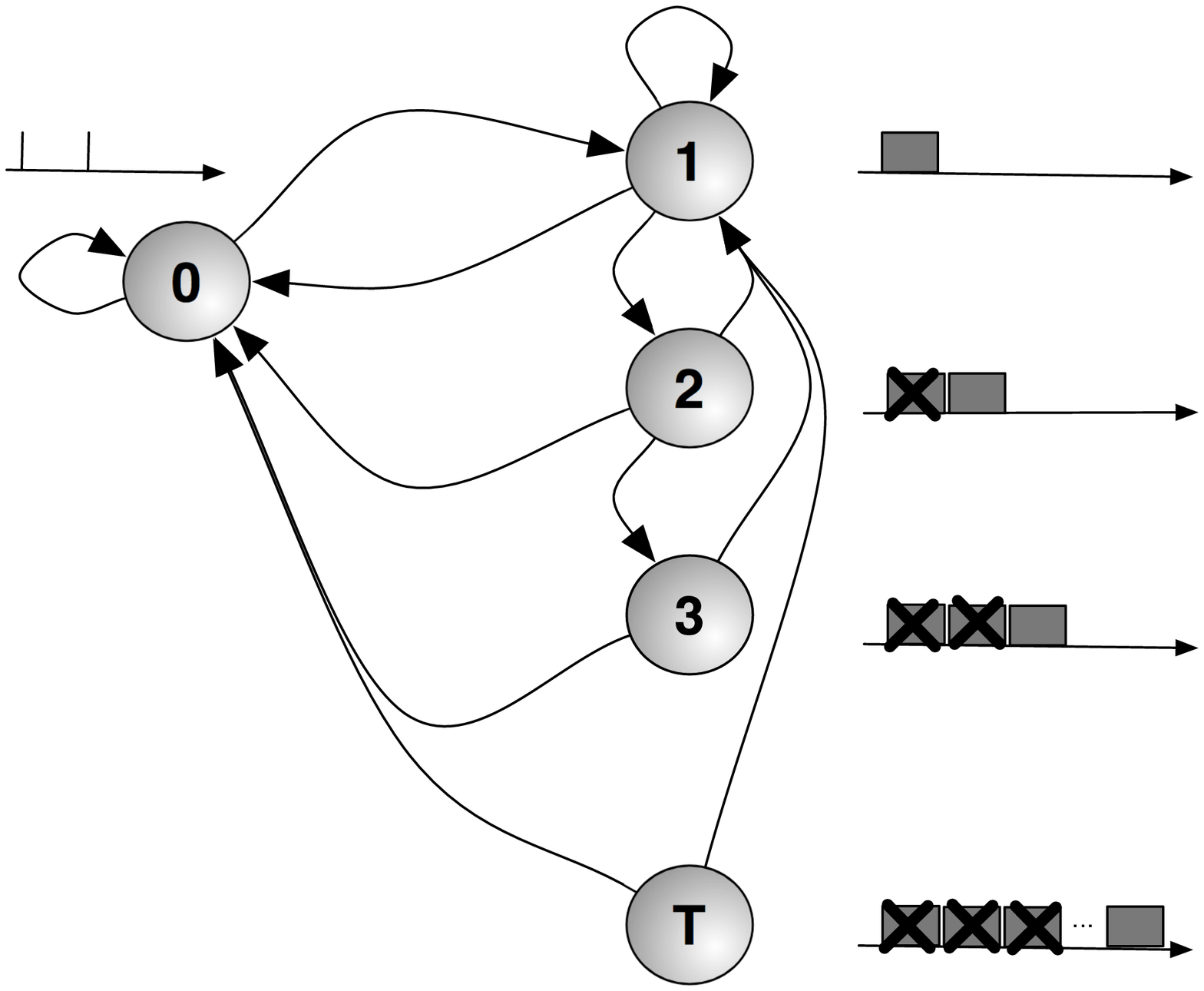}}
		\vup
	\caption{a) Primary source's state scheme. b) Graphical representation of the Markov chain of the system.}
\end{figure*}

Consider the network in Fig.~\ref{netfig}, with a primary and a secondary source, namely $S_{\rm P}$ and $S_{\rm S}$.
The primary source $S_{\rm P}$ and the secondary source $S_{\rm S}$ transmit packets to their respective destinations, namely
$D_{\rm P}$ and $D_{\rm S}$.

The reception of a packet at a particular destination is interfered with by the transmission of the other source. Our model subsumes the white space approach which typically assumes that a collision results in a decoding failure. An alternative view is that the collision approach implies a secondary access policy that will result in no throughput loss for the primary user. In contrast, we assign decoding error probabilities to the primary and secondary destinations as an abstraction of various interference mitigation methods. This in turn will result in some throughput loss and increase of the packet
failure probability as a function of the access strategy of the secondary user. It is trivial to show that the white space approach is optimal for the constraint of no collisions.

We assume a quasi-static channel model, where time is divided into slots of fixed duration and the channel gain of a certain link remains
constant within a slot, and is independent of the channel gains in the other slots. We denote by $g_{\rm PP}$, $g_{\rm PS}$,
$g_{SS}$ and $g_{\rm SP}$, the random variables corresponding to the channel coefficients respectively between $S_{\rm P}$ and
$D_{\rm P}$, $S_{\rm P}$ and $D_{\rm S}$, $S_{\rm S}$ and $D_{\rm S}$ and $S_{\rm S}$ and $D_{\rm P}$, and 
with $\zeta_{\rm PP}(g)$, $\zeta_{\rm PS}(g)$, $\zeta_{\rm SS}(g)$ and $\zeta_{\rm SP}(g)$ their respective probability
density function.

Assuming that the transmission of
a packet fits a slot, the performance of the receiver can be modeled via the average decoding failure probability, that depends on 
the packet encoding, transmission rates, structure of the receiver and average channel gains, as well as the activity of the concurrent source.
The average decoding failure probability at the primary destination $D_{\rm P}$ associated with a silent secondary source is denoted by $\rho{>}0$, while the same probability when the secondary source transmits is $\rho^*{\geq}\rho$. Analogously, the average decoding failure probability at the secondary destination $D_{\rm S}$ when the primary source is silent and transmitting is denoted with $\nu{>}0$ and $\nu^*{\geq}\nu$, respectively.

The construction fits many models and assumptions on the architecture of the physical layer and the transmission protocols. For instance, one
may assume that the primary destination performs signal decoding unaware of the presence of the secondary
source, and thus treats its signal as noise, whereas the secondary receiver adopts a smarter decoding strategy, by either treating
as noise or decoding and canceling the signal from the primary source according to the transmission rates, powers and channel
coefficients~\cite{covbook}.

Denoting the transmission rate and power of the primary and secondary sourcse with $R_{\rm P}$, $P_{\rm P}$, $R_{\rm S}$, $P_{\rm S}$,\footnote{
In the following example, rates $R_{\rm P}$ and $R_{\rm S}$ are expressed in [bit/s/Hz] and the transmission powers $P_{\rm P}$ and $P_{\rm S}$ are normalized to the noise power.}
respectively, we obtain the following failure probabilities for the primary link
\begin{eqnarray}
\rho\!\!&\!\!\!{=}\!\!\!&\!\!\mathcal{P}\left\{R_{\rm P}{>}\mathcal{C}\left(g_{\rm PP}P_{\rm P}\right)\right\}\\
\rho^{*}\!\!&\!\!\!{=}\!\!\!&\!\!\mathcal{P}\left\{R_{\rm P}{>}\mathcal{C}\left(\frac{g_{\rm PP}P_{\rm P}}{1{+}g_{\rm SP}P_{\rm S}}\right)\right\}
\end{eqnarray}
where $\mathcal{C}(x){=}\log (1{+}x)$.

For the secondary link we obtain
\begin{eqnarray}
\nu\!\!&\!\!\!=\!\!\!&\!\!\mathcal{P}\left\{R_{\rm S}{>}\mathcal{C}\left(g_{\rm SS}P_{\rm S}\right)\right\}\\
\nu^{*}\!\!&\!\!\!=\!\!\!&\!\!\mathcal{P}\left\{\{R_{\rm P},R_{\rm S}\}{\notin}\Psi\right\},
\end{eqnarray}
where $\Psi$ is the set of all the rate pairs $\{R_{\rm P},R_{\rm S}\}$ such that
\begin{eqnarray}
\label{set1}
R_{\rm S}\!\!&\!\!\!\leq\!\!\!&\!\!\mathcal{C}\left(g_{\rm SS}P_{\rm S}\right)\\
\label{set1bis}
R_{\rm P}+R_{\rm S}\!\!&\!\!\!\leq\!\!\!&\!\!\mathcal{C}\left(g_{\rm PP}P_{\rm P}+g_{\rm SS}P_{\rm S}\right),
\end{eqnarray}
or
\begin{equation}
\label{set2}
R_{\rm S}\leq\mathcal{C}\left(\frac{g_{\rm SS}P_{\rm S}}{1{+}g_{\rm PS}P_{\rm P}}\right),
\end{equation}
where Eqs.~(\ref{set1}) and~(\ref{set1bis}) refer to the achievable rate region corresponding to the secondary receiver
performing interference cancellation, while Eq.~(\ref{set2}) refers to the case in which the signal from the primary source
is treated as noise by the secondary receiver.
The failure probabilities listed above admit a simple integral form and can be easily computed. 

We remark that we do not consider a specific physical layer architecture or transmission technique, but rather, we refer
to the simple construction based on the average decoding probabilities described before.

In order to improve reliability, the primary source implements a retransmission-based error control scheme, by which a failed
packet is retransmitted in the subsequent slot. We consider a finite-retransmission process, where each packet can be transmitted
at most $T$ times, see Fig.~\ref{harq_sch}. Delayed retransmissions do not alter the following discussion.
If the packet has been transmitted $T$ times, it is discarded by the primary source. It is assumed that the destination sends an acknowledgment packet after each received packet, in order to make the source aware of the outcome of the transmission. Note that this scheme can be classified either as an automatic retransmission request (ARQ) or a type-I hybrid ARQ scheme depending on whether or not
the packets are encoded before transmission.
For the sake of simplicity, the secondary source is assumed to transmit each packet only once. This assumption is consistent with the
common characterization of secondary users as opportunistic sources without strict quality of service guarantees (``best effort''). 

Unless a retransmission is scheduled, the primary source $S_{\rm P}$ accesses the channel in each slot to
transmit a fresh packet with fixed probability $\alpha$, with $0{<}\alpha{<}1$. 
The secondary source is assumed to be backlogged, \emph{i.e.}, it always has a packet to transmit. However, a packet arrival
process at the secondary source can be included in the model with some straightforward modifications to
the analysis of the following section. Nevertheless, its inclusion does not add any insight to the discussion presented in this paper,
while it complicates the formulae. 

The channel access strategy of the secondary source follows a policy $\mu$, whose action set is $U={0,1}$, where $0$
and $1$ correspond to a silent and a transmitting source, respectively. We remark that transmission by the 
secondary source increases the probability of decoding failure at the primary receiver. Thus, the transition 
probabilities of the Markov chain strongly depend on the secondary user's activity and there is an explicit dependence 
between the stochastic characterization of the primary source activity and the activity of the secondary source.

The following discussion is specialized to a constraint defined on the throughput loss of the primary source. A constraint posed
on the increase of the failure probability only results in a different definition of the average primary source's cost,
as reported in Section~\ref{avfprob}.

The throughput achieved by the primary source under policy $\mu$ can be written as
\begin{equation}
\label{timeav}
\mathcal{W}_{\rm P}(\mu) {=}\lim_{N\rightarrow+\infty}\sup \frac{1}{\tau N}\sum_{n=1}^{N} \mathbb{E}[\mathit{I}(\Xi^{n}_{\rm P}(\mu))] L_{\rm P},
\end{equation}
where $L_{\rm P}$ is the size, in bits, of the packets sent by the primary source, $\tau$ is the duration of a slot, $\Xi^{n}_{\rm P}(\mu)$
is the event corresponding to a successfully delivered packet by the primary source in slot $n$, $\mathit{I}$ is the indicator function
and $\mathbb{E}$ denotes average.
The throughput of the secondary user admits an analogous expression.

The goal of the secondary source is to maximize its own achieved throughput while limiting throughput loss
to the primary source. In particular, let us denote as $\mu_0$ the policy by which the secondary source never transmits.
The optimization problem can be written as the following infinite horizon constrained Markov decision process~\cite{bert2}:
\begin{equation}
\label{optprob}
\widehat{\mu} {=} \arg\min_{\mu} \mathcal{J}_{\rm S}(\mu)~~{\rm s.t.}~~  \mathcal{W}_{\rm P}(\mu_0){-}\mathcal{W}_{\rm P}(\mu){\leq} \sigma,
\end{equation}
where $\mathcal{J}_{\rm S}(\mu)$ is the average cost incurred by the secondary source\footnote{In the following we will denote by
$\mathcal{J}_{\rm P}$ the analogous cost defined for the primary source.} and can be computed as 
$\mathcal{J}_{\rm S}(\mu)=L_{\rm S}/\tau{-}\mathcal{W}_{\rm S}(\mu)$. For two arbitrary policies $\mu_1$ and $\mu_2$, we refer to $\mathcal{W}_{\rm P}(\mu_1){-}\mathcal{W}_{\rm P}(\mu_2)$ as $\Delta(\mu_1,\mu_2)$. Note that the throughput loss can be also defined as the difference of average costs $\Delta(\mu_1,\mu_2){=}\mathcal{J}_{\rm P}(\mu_2){-}\mathcal{J}_{\rm P}(\mu_1)$.

According to~\cite{ross_constr}, the solution of the optimization problem (\ref{optprob}) is a past-independent randomized policy.
Moreover, as the number of independent constraints is equal to one, then, in the optimal stationary policy,
randomization occurs in at most one state, \emph{i.e.}, the map is either deterministic in all states or deterministic in all states
except one in which the decision is randomized. 

\section{Markov Chain and Optimization of the Network}
\label{opt}

The state of the network can be modeled as a homogeneous Markov process $\boldsymbol{\Theta}{=}\{\Theta_1,\Theta_2,\ldots\}$ taking values in the state space $\mathcal{X}{=}\{0,1,\ldots,T\}$, where $\Theta_n{=}0$ and $\Theta_n{=}\theta$, $1{\leq}\theta{\leq}T$, correspond to $S_{\rm P}$ not accessing the channel and performing the $\theta$--th transmission of a packet in slot $n$, respectively. Since the secondary source is backlogged and transmits each packet only once, its status is the same in each slot and we do not need to account for it in the model.
A graphical representation of the Markov chain is depicted in Fig.~\ref{cat}.

It can be shown that the solution of the problem in~(\ref{optprob}) is a randomized past-independent stationary policy~\cite{ross_constr}. Thus, the policy $\mu$
maps the state of the network $\theta{\in}\mathcal{X}$ to the probability that the secondary source takes the actions
in $\mathcal{U}$. The action selected in the time slot $n$ is referred to as $u_n{\in}\mathcal{U}$ in the following. We define $\mu(\theta,u)$ as the probability that the secondary source takes action $u$ when the network is in state $\theta$.
As $\mathcal{U}$ is binary, the policy can be defined as the vector $\underline{\kappa}{=}\{\kappa_0,\kappa_1,\ldots,\kappa_{T}\}$,
where $0{\leq}\kappa_{\theta}{=}\mu(\theta,1){\leq}1$. $\kappa_{\theta}$ is the probability that the secondary source accesses the channel when the network
is in state $\theta$. Policy $\mu_0$ corresponds to the all-zero vector $\underline{\mathit{0}}$. For the sake of simplicity, in the following
$L_{\rm P}/\tau{=}L_{\rm S}/\tau$ is set to one.

The transition probability from state $\Theta$ to state $\Theta^{\prime}$, $\Theta,\Theta^{\prime}{\in}\mathcal{X}$, conditioned on the
action $u$ is defined as
\begin{equation}
\zeta_u(\Theta,\Theta^{\prime}){=}\mathcal{P}\left\{\Theta_{n+1}=\Theta^{\prime}|\Theta_{n}=\Theta,u_n{=}u\right\},
\end{equation}
and does not depend on $n$.
Note that since the policy $\mu$ is past-independent, then the stochastic process which models the temporal evolution of the network is a
Markov process. We remark that due to the mutual interference the probability that the Markov process transitions from one state to 
another in the state space depends on the action taken of the secondary user.

The transition matrix of the chain, which collects the transition probabilities $\zeta^{\mu}(\Theta,\Theta^{\prime})$, is
\begin{equation}
\begin{bmatrix}
\label{matprim}
1{-}\alpha  &  \alpha & 0 & \ldots &0 \\
(1{-}\alpha)(1{-}\rho_1)  &  \alpha(1{-}\rho_1) & \rho_1 & \ldots &  0\\
\vdots & \vdots & \vdots & \ddots & \vdots\\
(1{-}\alpha)(1{-}\rho_{T-1})  &  \alpha(1{-}\rho_{T-1}) & 0 & \ldots &\rho_{T-1}\\
(1{-}\alpha)  &  \alpha & 0 & \ldots &0
\end{bmatrix},
\end{equation}
where $\rho_\theta$ depends on the transmission probability of the secondary user $\kappa_\theta$. and represents the failure probability of the primary source in state $t$ conditioned on the transmission strategy. Thus, from state $0$,
the network moves to state $1$ (new packet in the buffer of $S_{\rm P}$) with probability $\alpha$, and remains in $0$ otherwise. In each state ${\theta}$, $1{\leq}{\theta}{<}T$,
the network moves to state ${\theta}{+}1$ if a failure occurs, while it returns to $0$ or $1$ according to the arrival probability $\alpha$
if the primary packet is successfully delivered. From state $T$, the transmission of the current packet is terminated regardless
of failure or success, and thus the network returns to state $0$ and $1$ with probability $1{-}\alpha$ and $\alpha$, respectively.

The steady-state distribution $\pi_{\mu}$ of the Markov chain is the solution of the following system of equations
\begin{eqnarray}
\label{pisis}
\pi_{\mu}(0) \!&\!\!\!{=}\!\!\!&\!\! (1{-}\alpha)\pi_{\mu}(0) {+} (1{-}\alpha)\!\!\sum_{t=1}^{T-1}(1{-}\rho_t)\pi_{\mu}(t){+}(1{-}\alpha)\pi_{\mu}(T)\nonumber\\
\pi_{\mu}({\theta}) \!&\!\!\!{=}\!\!\!&\! \rho_{{\theta}}\pi_{\mu}({{\theta} -1}) ~~~~~{\rm for}~~2{\leq}{\theta}{\leq}T,
\end{eqnarray}
with the normalization condition $\pi_{\mu}(1){=} 1-\pi_{\mu}(0) -\sum_{t=2}^{T}\pi_{\mu}(t)$.

As intuition suggests, states corresponding to a larger number of transmissions are hit by the process a smaller number of times
with respect to those associated with a smaller number of transmissions of the same packet, \emph{i.e.}, $\pi_{\mu}({\theta}{+}1){\leq}\pi_{\mu}({\theta})$ for any ${\theta}{>}0$. In fact, the process enters state ${\theta}{+}1$, ${\theta}{>}0$, only by passing through ${\theta}$. This can be observed in Eq.~(\ref{pisis}), by which we get
$\pi_{\mu}({\theta}) {=} \pi_{\mu}(1)\prod_{i=1}^{{\theta}-1}\rho_i$, for $2{\leq}{\theta}{\leq}T$,
where $\prod_{i=1}^{{\theta}-1}\rho_i{\leq}1$.
The steady-state distribution is
\begin{eqnarray}
\label{stdistr}
\pi_{\mu}(0) \!&\!\!\!{=}\!\!\!&\! \frac{1-\alpha}{1+\alpha\sum_{t=1}^{T-1}\prod_{i=1}^{t}\rho_i},\nonumber\\
\pi_{\mu}(1) \!&\!\!\!{=}\!\!\!&\! \frac{\alpha}{1+\alpha\sum_{t=1}^{T-1}\prod_{i=1}^{t}\rho_i},\nonumber\\
\pi_{\mu}(\theta) \!&\!\!\!{=}\!\!\!&\! \frac{\alpha\prod_{i=1}^{{\theta}-1}\rho_i}{1+\alpha\sum_{t=1}^{T-1}\prod_{i=1}^{t}\rho_i}~~~~{\rm for}~~2{\leq}t{\leq}T.
\end{eqnarray}
The average cost of the primary source can be rewritten as
$\mathcal{J}_{\rm P}(\mu){=}\sum_{\theta{\in}\mathcal{X}}\pi_{\mu}({\theta})\tilde{\gamma}_{\rm P}(\mu,{\theta})$,
where $\tilde{\gamma}_{\rm P}(\mu,{\theta})$ is the average cost collected by the primary source in state $\theta$ under policy $\mu$.

The average cost difference $\Delta(\mu_1,\mu_2)$ is equal to
\begin{equation}
\Delta(\mu_1,\mu_2){=}\sum_{{\theta}{\in}\mathcal{X}}(\pi_{\mu_2}({\theta})\tilde{\gamma}_{\rm P}(\mu_2,\theta){-}\pi_{\mu_1}({\theta})\tilde{\gamma}_{\rm P}(\mu_1,{\theta})).
\end{equation}
Different policies result in different average costs collected in each state, but correspond to different steady-state distributions as well.
\begin{figure*}[!b]
\hrulefill
\vspace*{1pt}
\normalsize
\setcounter{mytempeqncnt}{\value{equation}}
\setcounter{equation}{17}
\begin{equation}
\label{thr_sec}
\mathcal{W}_{\rm S}\left(\underline{\kappa}\right)=\pi_{\underline{\kappa}}(0)\kappa_0(1-\nu)+\!\!\! \sum_{\theta=0}^{T}\!\!\pi_{\underline{\kappa}}(\theta)\kappa_{\theta}(1-\nu^{*})=\frac{(1{-}\alpha)(1-\nu){+}(1-\nu^{*})\lambda_S\alpha\sum_{t=1}^{T}\kappa_{t}\prod_{i=1}^{t-1}(\rho{+} (1 {-} \rho)\lambda \kappa_i))}{1{+}\alpha\sum_{t=1}^{T-1}\prod_{i=1}^{t}(\rho{+} (1 {-} \rho)\lambda \kappa_i)}.
\end{equation}
\setcounter{equation}{\value{mytempeqncnt}}

\vup
\end{figure*}

The average cost in state $\theta$ can be computed as
\begin{equation}
\tilde{\gamma}_{\rm P}(\mu,\theta){=}\sum_{\theta_1\in\mathcal{X}}\sum_{u\in\mathcal{U}}\mu(\theta,u)\gamma_{\rm P}(\theta,\theta_1)\zeta_u(\theta,\theta_1),
\end{equation}
where $\gamma_{\rm P}(\theta,\theta_1)$ is the cost incurred by $S_{\rm P}$ during the transition from $\theta$ to $\theta_1$. The cost $\gamma_{\rm P}(\theta,\theta_1)$ is equal to zero for the transitions in which a packet is successfully delivered and to $L_{\rm P}/\tau{=}1$ when a packet incurs failure.\footnote{We recall that, without any loss of generality, $L_{\rm P}/\tau$ is set to
unity.} Note that when $\theta{=}T$, the cost of any transition is $\rho_T L_{\rm P}/\tau{=}\rho_T$. 
The throughput can be similarly defined as the sum of the steady-state distribution weighed by the average rewards $\tilde{\omega}(\mu,\theta){=}
L_{\rm P}/\tau{-}\tilde{\gamma}_{\rm P}(\mu,\theta){=}1{-}\tilde{\gamma}_{\rm P}(\mu,\theta)$. Analogous definitions can be stated for the secondary link. In the following, with a slight 
abuse of notation, we denote the average cost and reward from state $\theta$ when the action $u$ is selected as $\tilde{\gamma}(u,\theta)$, $\tilde{\omega}(u,\theta)$, respectively.

The average costs in the various states are trivially $\tilde{\gamma}_{\rm P}(\mu,0){=}L_{\rm P}/\tau{=}1$ and $\tilde{\gamma}_{\rm P}(\mu,{\theta}){=}\rho_{\theta}L_{\rm P}/\tau{=}\rho_{\theta}$, $1{\leq}{\theta}{\leq}T$. The average cost of the primary source can be thus written as
\begin{equation}
\mathcal{J}_{\rm P}\left(\underline{\rho}\right){=}\frac{(1{-}\alpha){+}\alpha\sum_{t=1}^{T}\prod_{i=1}^{t}\rho_i}{1{+}\alpha\sum_{t=1}^{T-1}\prod_{i=1}^{t}\rho_i}{\leq}1,
\end{equation}
with $\underline{\rho}{=}\{\rho_1,\ldots,\rho_{T}\}$.

The interference by the secondary source in a certain state has two effects on the performance of the primary source:
\begin{itemize}
\item if ${\theta}{>}0$, interference increases the instantaneous cost collected in that state by the primary source;
\item if $0{<}{\theta}{<}T$, interference increases the probability that the process moves to $\theta{+}1$.
\end{itemize}
Clearly, transmission by $S_{\rm S}$ in state $0$ does not have any effect on
the primary source, while in $T$ it only increases the cost associated with that state, as the packet being served by $S_{\rm P}$ is discarded after this transmission.

As observed in the Introduction, if the secondary source transmits in a state ${\theta}$, with $0{<}{\theta}{<}T$, the average number of
transmissions of the packets of the primary source increases. This means that the fraction of time spent by the primary source
in the idle state decreases. By interfering with the primary source, the secondary source is then decreasing the number of idle
slots, that is, the white spaces reduce.

The average failure probability of the primary source in state ${\theta}{>}0$, conditioned on the policy, is $\rho_{\theta}{=}(1{-}\kappa_{\theta})\rho {+} \kappa_{\theta} \rho^*$. In fact, when in state $\theta$, the primary source incurs a failure probability equal to $\rho^*$ if the secondary source transmits and equal to $\rho$ if the secondary source does not transmit. 
\begin{figure}[t]
	\centering
	\includegraphics[width=0.4\columnwidth]{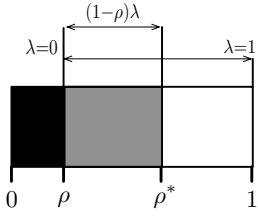}
\caption{Graphical representation of the failure probability increasing factor associated with the interference generated by the secondary source to the primary source's transmission.}
\vup
\label{lam}
\end{figure}

In order to provide a more intuitive explanation of the dependence between the decoding performance degradation at $D_{\rm P}$ and transmission
by $S_{\rm S}$, we define the \emph{failure probability increasing factor} $\lambda$, such that $\rho^*{=}\rho{+}(1{-}\rho)\lambda$. 
Thus, $\lambda$ determines the impact of transmission by the secondary source on the decoding probability at the primary receiver: the
larger $\lambda$, the closer to one the probability of failure. In particular, for $\lambda{=}0$ and $\lambda{=}1$, the failure probability at
the primary source is $\rho$ and $1$, respectively (see Fig.~\ref{lam} for a graphical representation). The resulting average failure rate in ${\theta}$ is $\rho_{\theta}{=}\rho{+} (1 {-} \rho)\lambda \kappa_{\theta}$. Note that $\lambda$ parameterizes the difference between the failure probability with and without interference from the secondary user and does not presume the use of a linear model for the failure probability as a function of the interference power.

Consider $\kappa_0$, the probability that the secondary source transmits when the primary source is idle. As increasing $\kappa_0$ does not affect the cost to the primary user (which is idle), the optimal value for $\kappa_0$ is one.\footnote{This may not hold if we consider more complex networks or energy consumption metrics. This is left for future research.} Thus, in the sequel, we set $\kappa_0{=}1$.

The average cost collected by the primary source is then
\begin{equation}
\mathcal{J}_{\rm P}\left(\underline{\kappa}\right){=}\frac{(1{-}\alpha){+}\alpha\sum_{t=1}^{T}\prod_{i=1}^{t}(\rho{+} (1 {-} \rho)\lambda \kappa_i)}{1{+}\alpha\sum_{t=1}^{T-1}\prod_{i=1}^{t}(\rho{+} (1 {-} \rho)\lambda \kappa_i)}.
\end{equation}
The average throughput achieved by the secondary source, also referred to as the \emph{reward} in the following, can be computed as in Eq.~(\ref{thr_sec}).\addtocounter{equation}{1}

The optimization problem (\ref{optprob}) is equivalent to the following linear program (LP)~\cite{ross_constr}
\begin{align}
\label{LP}
\widehat{\boldsymbol{Z}}=\arg\max_{\boldsymbol{Z}}~& \sum_{\theta\in\mathcal{X}}\sum_{u\in\mathcal{U}}\tilde{\omega}(u,\theta)z_u(\theta)\\
{\rm s.t.}~
&\sum_{\theta\in\mathcal{X}}\sum_{u\in\mathcal{U}}\tilde{\gamma}(u,\theta)z_u(\theta){\leq}\sigma{+}\mathcal{J}_{\rm P}(\mu_0)\nonumber\\
&\sum_{\theta\in\mathcal{X}}\sum_{u\in\mathcal{U}}z_u(\theta){=}1\nonumber\\
&\sum_{u\in\mathcal{U}}z_u(\theta_1){=}\sum_{\theta\in\mathcal{X}}\sum_{u\in\mathcal{U}}z_u(\theta)\zeta_u(\theta,\theta_1), ~\forall \theta_1\nonumber\\
&z_u(\theta){\geq}0,~~ \forall u,~\theta,\nonumber
\end{align}
where $\boldsymbol{Z}{=}\{z_{u}(\theta)\}_{\theta\in\mathcal{X},u\in\mathcal{U}}$, 
and $z_u(\theta)$ represents the joint probability that the Markov chain is in state $\theta$ and action $u$ is selected. 
The first constraint bounds the maximum performance loss of the primary user, while the others force the solution to be a valid stationary distribution
for the Markov chain.

The LP defined above thus optimizes the steady-state distribution of state-action pairs. The involved expression of the average reward and cost
functions defining the objective and constraint of the original problem are thus translated into \emph{linear combinations} of the optimization variables.
\footnote{A constraint on the failure probability can be formalized as a linear constraint as well through straightforward manipulation.}
The condition for optimality is that the Markov chain under all
the policies is unichain \cite{MEYNbook}, \emph{i.e.}, it has a single recurrent class and an arbitrary number of transient classes. This property holds,
in our case, for any policy and any set of parameters as defined throughout the paper. 

The optimal policy is then $\widehat{\mu}(\theta,a){=}\widehat{z}_a(\theta)/(\sum_{u}\widehat{z}_u(\theta))$ if $\sum_{u\in\mathcal{U}}z_{u}(\theta)=1$,
\emph{i.e.}, $\theta$ is recurrent. If $\sum_{u\in\mathcal{U}}z_{u}(\theta)=0$, \emph{i.e.}, $\theta$ is transient, then the map in $\theta$ is
$\widehat{\mu}(\theta,a)=1$ for a randomly chosen $a\in\mathcal{U}$ and $\widehat{\mu}(\theta,a)=0$ otherwise.
In the model at hand, which considers a binary action whose randomization corresponds to the probability that the secondary source transmits
given that the transmission probability $\kappa_{\theta}$ simply corresponds to $\mu(\theta,1)$. Note that $\sum_{\theta=0}^Tz_{1}(\theta)$
is the total fraction of time in which the secondary source transmits.

It is also shown in~\cite{ross_constr} that the number of \emph{randomizations}, \emph{i.e.}, the number of states in which the policy
is non-deterministic, is equal to or smaller than the number of independent constraints in Equation~(\ref{LP}). Thus, in the model at hand, the optimal policy found via the above LP is non deterministic in at most one state and the optimal vector $\widehat{\underline{\kappa}}$ is a vector
with $N_{1}$ ones, $N_{0}$ zeros and $N_r$ elements in $(0,1)$, with $N_1{+}N_0{+}N_r{=}T{+}1$, $0{\leq}N_1{\leq}T$, $0{\leq}N_0{\leq}T$, and $N_r{=}1$ or $0$. The
space of the vectors described by the above conditions is denoted in the following with $\mathcal{M}_r$.

In the following Section, we will show that, if $\nu^*{=}\nu$, the optimal policy $\widehat{\underline{\kappa}}$ has a
precise structure that enables its calculation through a simple algorithm, thereby avoiding the need to solve the linear problem stated before.
In particular, the optimal policy concentrates transmissions by the secondary source in the first transmissions of the primary source packets.
Therefore, the $N_1$ unit elements and the $N_0$ zero elements are the first $N_1$ and the last $N_0$ elements of the vector $\widehat{\underline{\kappa}}$, respectively. If $N_1{+}N_0{=}T{-}1$, then randomization occurs at the $N_1{+}1$-th state,
otherwise the policy is deterministic.

\emph{As a side comment, we observe that in a pure collision scenario, where the failure a policy such that  $\kappa_0{=}1$ and
$\kappa_{\theta}{=}0$, $0{<}{\theta}{\leq}T$, is optimal}. This is the white spaces approach. In fact, if $\rho^{*}_{\rm P}$
 and $\rho^{*}_{\rm S}$ are both set to one,
the secondary source gains nothing when transmitting concurrently with the primary source, while increasing
the cost of the latter. In general, if the secondary source bases its strategy on channel sensing only it can distinguish
between an idle slot ($\theta{=}0$) and a non-idle slot ($0{<}\theta{\leq}T$). The resulting strategy assigns the transmission
probabilities $\kappa_0{=}1$ and $\kappa_{\theta}{=}\kappa$, $\forall 0{<}{\theta}{\leq}T$. We will show through numerical
results that this policy is suboptimal.

\section{Optimal Transmission Strategy for the Z-Interference Channel}
\label{optz}

In this Section, we address the structure of the optimal transmission strategy in the particular case in which $\nu^*{=}\nu$,
that is, the transmission by the primary source does not affect the successful decoding probability of the packet of the
secondary source by the secondary receiver.

This assumption can be referred to the well-known Z-interference channel framework, where the interference link between
the primary source and the secondary destination is removed. We observe that this does not mean that the interference
channel between the primary source and the secondary destination is simply removed. For instance, this model also fits the case
in which $g_{PS}{\gg}g_{\rm SS}$ with high probability, or the primary source transmits with a rate $R_{\rm P}$ sufficiently low to
allow the secondary destination to decode and cancel the interference from the primary source with high probability.

In this case $\nu^*{=}\nu$, and thus the failure probability at the secondary destination does not influence the solution of the optimization problem.
In fact, the success probability $1{-}\nu$ only represents a scaling factor for the reward achieved by the secondary source. Thus, in the
following, with $\mathcal{W}_{\rm S}(\underline{\kappa})$ we refer to the normalized reward $\mathcal{W}_{\rm S}(\underline{\kappa})/(1{-}\nu)$.
We remark that the optimal policy $\underline{\widehat{\kappa}}$ when maximizing the reward or the normalized reward of the
secondary source is the same, and that the throughput is simply the normalized reward multiplied by the success probability.

\subsection{Structure of the Optimal Policy}
In the following, we show that the optimal transmission policy for the secondary source when $\nu{=}\nu^*$ has a specific structure.
The transmission strategy maximizing the throughput of the secondary source, given the constraint on the primary
source's throughput loss, concentrates interference in the first transmissions of each of the packets sent by the
primary source. The policy has the structure described in Theorem~\ref{struct}, where the secondary user transmits with
probability $1$ in states $\theta{<}N_1$, probability $\kappa_{N_1}{\in}[0,1]$ in state $N_1$ and
probability equal to zero in states $\theta{>}N_1$. The values of $N_1$ and $\kappa_{N_1}$ are functions
of the parameters of the system and of the throughput constraint. It can be shown that the same structure applies if the constraint is
on the failure probability of primary source's packets. The definitions and proof for this last case are provided in Section~\ref{avfprob}.

As discussed before, interference from the secondary source in different states has a different effect. In fact,
if the secondary source increases its transmission probability in state $j$, with $0{<}j{<}T$, it also increases the average failure probability 
$\rho_j$. This means that the primary source fails more often in the $j$--th transmission of a packet. Therefore, the Markov process
hits more frequently the states with indices larger than $j$, and less frequently all the other states, that is, the steady-state probability 
$\pi_{\mu}(t)$ of the states $t{>}j$ grows, while the same probability associated with the states $t{\leq}j$ decreases.

Moreover, as observed before, if $j{<}r$, then the steady-state probability associated with state $j$ is larger than that
of state $r$. Thus, if the secondary source increases its transmission probability in state $j$, it increases the overall
level of interference more than if the same increase is applied to state $r$.
Thus, the state in which the interference is increased influences the bias on the stochastic process of the primary source,
due to the activity of the secondary source as well as the overall cost incurred by the former.
The normalized reward of the secondary source counts the fraction of slots in which the secondary source transmits.
Since $\pi_{\mu}(j){>}\pi_{\mu}(r)$, the overall reward grows more if the transmission probability is increased in state
$j$ than if the same increase is applied to state $r{>}j$. On the other hand, note that if $\kappa_j$ is increased, then $\pi_{\mu}(j)$
decreases. Nevertheless, we will shown in the following that, if $\nu^*{=}\nu$, an increased transmission probability in any of the states
results in a larger secondary source's throughput.

The main intuition behind the structure of the optimal transmission policy is that, when
considering the same increase of the transmission probability, \emph{the reward of the secondary source grows faster than
the cost of the primary source}. Moreover, the difference
between the increase of the reward of the secondary source and the increase of the cost of the primary source grows much faster if the
transmission probability is increased in state $j$, with respect to the same quantity measured if the transmission
probability is increased in state $r{>}j$. Based on these observations, it is possible to show that
among the set of the transmission policies resulting in the same cost for the primary source, the one that most concentrates
the interfering transmissions in the first transmissions of the primary source's packets achieves the optimal throughput.

We first state the following theorems:
\begin{theorem}
\label{th1}
$\mathcal{J}_{\rm P}(\underline{\kappa})$ is a strictly increasing function of $\kappa_{\theta}$, with $\theta{>}0$.\footnote{We remark that the cost is independent of $\kappa_0$, whose value has been set to one by assumption.}
\end{theorem}
\begin{theorem}
\label{th2}
$\mathcal{W}_{\rm S}(\underline{\kappa})$ is a strictly increasing function of $\kappa_\theta$, with $\theta{\geq}0$.
\end{theorem}
Formal proofs of these Theorems are provided in Appendices~\ref{app1} and~\ref{app2}.

Theorem~\ref{th1} states that the cost of the primary source increases as the fraction of slots in which the secondary user accesses the channel
gets larger. This is rather intuitive, as a larger amount of interference cannot result in a larger throughput for the interfered link, at least in the
framework considered herein. 

Theorem~\ref{th2} states that the average throughput of the secondary source increases as the fraction of slots in which it accesses the
channel gets larger. Although this result also agrees with intuition, it must be observed that transmission by the secondary source in a
certain slot also modifies the steady-state distribution of the Markov chain for the primary source. For instance, the steady-state distribution of state $0$, in which the secondary source
can always transmit, decreases as $\kappa_{\theta}$ gets larger, with $0{<}{\theta}{<}T$.\footnote{In the case ${\theta}{=}T$, transmission by the secondary source
does not modify the steady-state distribution.} However, the theorem states that the gain outweighs the potential loss under the assumptions on the decoding failure at the
secondary receiver stated before.

The previously stated theorems guarantee that the optimal policy lies in the space of policies where the constraint on the primary throughput loss
of Eq.~(\ref{optprob}) is active, \emph{i.e.}, $\Delta(\underline{\mathit{0}},\widehat{\underline{\kappa}}){=}\sigma$, unless $\Delta(\underline{\mathit{0}},\underline{\mathit{1}}){\leq}\sigma$, where $\underline{\mathit{1}}$ is a $T{+}1$--long vector whose elements are all ones.
In fact, in this latter case, the secondary source transmits in all the slots with probability one, and, if this policy results in a cost
for the primary source smaller than the maximum admitted, then a policy that activates the constraint does not exist. Moreover,
under the policy $\underline{\mathit{1}}$, the secondary user achieves the maximum possible throughput, \emph{i.e.},
$\mathcal{W}_{\rm S}(\underline{\mathit{1}}){=}L_{\rm S}/\tau{=}1$.\footnote{We recall that, throughout this section, we normalize
the throughput of the secondary source normalized to the success probability $1{-}\nu{=}1{-}\nu^*$.} Thus, if $\underline{\mathit{1}}$ is admissible, then it is also optimal.

Let us consider now the case $\Delta(\underline{\mathit{0}},\underline{\mathit{1}}){>}\sigma$ and define $\underline{u}_i$ as a $T{+}1$-long vector of all zeros except for the $i$--th element that is equal to one, $0{\leq}i{\leq}T$. 

Consider a policy $\underline{\kappa}^{\prime}{\neq}\underline{\mathit{1}}$ such that $\Delta(\underline{\mathit{0}},\underline{\kappa}^{\prime}){<}\sigma$.
Since $\mathcal{J}_{\rm P}$ is continuous, there exist $\delta{>}0$ and $0{\leq}j{\leq}T$ such that $\Delta(\underline{\mathit{0}},\underline{\kappa}^{\prime\prime}){\leq}\sigma$, where $\underline{\kappa}^{\prime\prime}{=}\underline{\kappa}^{\prime}{+}\underline{u}_j\delta$. 
Due to Theorem~\ref{th2}, the reward achieved by policy $\underline{\kappa}^{\prime\prime}$ is larger than that achieved by policy $\underline{\kappa}^{\prime}$, \emph{i.e.}, $\mathcal{W}_{\rm S}(\underline{\kappa}^{\prime\prime}){>}\mathcal{W}_{\rm S}(\underline{\kappa}^{\prime})$.
Note that for any $\delta{>}0$, we also have $\mathcal{J}_{\rm P}(\underline{\kappa}^{\prime\prime}){>}\mathcal{J}_{\rm P}(\underline{\kappa}^{\prime})$, and thus $\sigma{-}\Delta(\underline{\mathit{0}},\underline{\kappa}^{\prime\prime}){<}\sigma{-}\Delta(\underline{\mathit{0}},\underline{\kappa}^{\prime})$.

Thus, for any policy $\underline{\kappa}^{\prime}$ resulting in a maximum performance loss below $\sigma$, there exists an
admissible policy $\underline{\kappa}^{\prime\prime}$ such that the secondary user achieves an improved throughput, while the cost
of the primary user increases. Any policy $\underline{\kappa}^{\prime}$ such that 
$\Delta(\underline{\mathit{0}},\underline{\kappa}^{\prime})$ is strictly smaller than $\sigma$ is thus non-optimal.

As a consequence of the previous statements, if the problem in~(\ref{optprob}) is feasible, then the optimal policy $\widehat{\underline{\kappa}}$ lies in the space $\mathcal{M}_{\sigma}{=}\{\underline{\kappa}{:}\Delta(\underline{\mathit{0}},\underline{\kappa}){=}\sigma\}{\cup}\{\underline{\mathit{1}}\}$.

\begin{figure}[t]
	\centering
	\includegraphics[width=0.7\columnwidth]{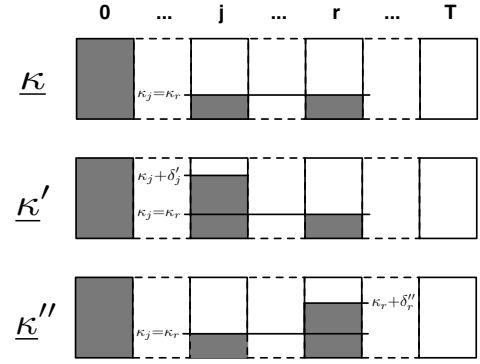}
\caption{Policies $\underline{\kappa}^{\prime}$ and $\underline{\kappa}^{\prime\prime}$ as defined in Theorems~\ref{mainth}
and~\ref{mainth2}}
\vup
\label{mainthfig}
\end{figure}

We now formalize the intuition discussed before by stating the following theorem:
\begin{theorem}
\label{mainth}
Consider a policy $\underline{\kappa}$ such that $\kappa_j{=}\kappa_r$ and $\kappa_{\theta}{=}0$, $\forall {\theta}{>}r$ and with $0{<}j{<}r{\leq} T$.

Define the two policies $\underline{\kappa}^{\prime}$ and $\underline{\kappa}^{\prime\prime}$ as $\underline{\kappa}^{\prime}{=}\underline{\kappa}{+}\underline{u}_j\delta_j^{\prime}$ and $\underline{\kappa}^{\prime\prime}{=}\underline{\kappa}{+}\underline{u}_r\delta_r^{\prime\prime}$, with $0{<}\delta_j^{\prime}{\leq}1{-}\kappa_j$ and $0{<}\delta_r^{\prime\prime}{\leq}1{-}\kappa_r$ (see Fig.~\ref{mainthfig} for a graphical representation).
If $\mathcal{J}_{\rm P}(\underline{\kappa}^{\prime}){=}\mathcal{J}_{\rm P}(\underline{\kappa}^{\prime\prime})$ then $\mathcal{W}_{\rm S}(\underline{\kappa}^{\prime}){>}\mathcal{W}_{\rm S}(\underline{\kappa}^{\prime\prime})$.
\end{theorem}
The proof of the theorem is provided in Appendix~\ref{appmth}.

Theorem~\ref{mainth} states that, starting from a policy $\underline{\kappa}$ respecting the hypothesis, if the policy obtained by increasing
$\kappa_j$ and the policy obtained by increasing $\kappa_r$
incur the same average primary source's cost, then,
if $j{<}r$, the reward associated with the former is larger than the reward associated with the latter.
As discussed before, this result is due to the difference between
the reward and cost increase corresponding to an increased transmission probability in a certain state. This
quantity grows faster if the transmission probability is increased in state $j$ with respect to state $r$, with $j{<}r$.

Similarly, it can be shown that if the policy obtained by decreasing $\kappa_j$ and the policy
obtained by decreasing $\kappa_r$ result in the same
average primary source's cost, then the reward achieved by the latter is larger than the reward achieved by the former.
Formally:
\begin{theorem}
\label{mainth2}
Consider a policy $\underline{\kappa}$ such that $\kappa_j{=}\kappa_r$ and $\kappa_{\theta}{=}0$, $\forall {\theta}{>}r$ and with $0{<}j{<}r{\leq} T$.

Define the two policies $\underline{\kappa}^{\prime}$ and $\underline{\kappa}^{\prime\prime}$ as $\underline{\kappa}^{\prime}{=}\underline{\kappa}{-}\underline{u}_j\delta_j^{\prime}$ and $\underline{\kappa}^{\prime\prime}{=}\underline{\kappa}{-}\underline{u}_r\delta_r^{\prime\prime}$, with $0{<}\delta_j^{\prime}{\leq}\kappa_j$ and $0{<}\delta_r^{\prime\prime}{\leq}\kappa_r$ (see Fig.~\ref{mainthfig} for a graphical representation).
If $\mathcal{J}_{\rm P}(\underline{\kappa}^{\prime}){=}\mathcal{J}_{\rm P}(\underline{\kappa}^{\prime\prime})$ then $\mathcal{W}_{\rm S}(\underline{\kappa}^{\prime}){<}\mathcal{W}_{\rm S}(\underline{\kappa}^{\prime\prime})$.
\end{theorem}
The proof of the theorem is provided in Appendix~\ref{appmth2}.

Theorem~\ref{mainth} and~\ref{mainth2} are the basis for the derivation of the structure of the optimal policy $\widehat{\underline{\kappa}}$,
defined by the following theorem:
\begin{theorem}
\label{struct}
The optimal policy $\widehat{\underline{\kappa}}$ has the following structure
\begin{equation}
\widehat{\underline{\kappa}}{=}[\underline{\mathit{1}}_{N_1},~ \kappa_{N_1},~ \underline{\mathit{0}}_{N_0}],
\end{equation}
where $\underline{\mathit{1}}_{N_1}$ and $\underline{\mathit{0}}_{N_0}$ are vectors of $N_1$ ones and $N_0$ zeros,
respectively, and $0{\leq}\kappa_{N_1}{\leq}1$.
\end{theorem}

Thus, the optimal policy concentrates transmission by the secondary source in those states associated with the first
transmissions of primary source's packets. 
Intuitively, if the interference generated by the primary user's transmission has a small impact on the reception of secondary user's packets, then the 
difference between the secondary user reward increase and the primary user cost increase corresponding to an increase of the 
transmission probability in the early retransmissions is positive and larger than that corresponding to the same increase in the late retransmissions.
In fact, the throughput achieved by the secondary user is not affected by the access rate of the primary user, and thus, a transmission probability increase
corresponds to a positive reward in all the states. Moreover, the cost increase of the primary user accounts for the fact that additional primary user's retransmission
due to secondary user interference take place in otherwise idle slots with a positive probability, that is, there is a positive probability that
retransmissions do not affect the throughput of the primary user. This reduces the cost increase speed in the early retransmissions of primary user
packets and results into the unique structure of the optimal transmission policy discussed before.

We remark that the optimal transmission strategy of the secondary user is defined under the constraint on the maximum performance
loss of the primary user. Therefore, the transmission probability in all the states $\Theta{>}0$ is bounded by the constraint. We also observe
that the transmission strategies proposed in prior literature addressing cognitive networks do not consider the long term impact of interference.
Therefore, these strategies may fail to guarantee the minimum performance to the primary user in those scenarios in which the primary user
implements protocols and mechanisms which react to interference and packet failure.

Theorem~\ref{struct} has a very intuitive proof, sketched in the following. As observed before, if the problem in Eq.~(\ref{optprob}) is feasible,
then the policy lies in the space of policies $\mathcal{M}_{\sigma}{=}\{\underline{\kappa}{:}\Delta(\underline{\mathit{0}},\kappa){=}\sigma\}{\cup}\{\underline{\mathit{1}}\}$.
Moreover, according to~\cite{ross_constr}, the optimal policy is a randomized policy with randomization in at most one state.
We recall that the space of transmission probability vectors associated with those policies,
\emph{i.e.}, the space of the vectors with $N_{1}$ ones, $N_{0}$ zeros and $N_r$
elements in $(0,1)$, with $N_1{+}N_0{+}N_r{=}T{+}1$, $0{\leq}N_1{\leq}T$, $0{\leq}N_0{\leq}T$, and $N_r{=}1$ or $0$,
is denoted with $\mathcal{M}_r$. Therefore, the optimal transmission probability vector lies in the space
$\mathcal{M}_r{\cap}\mathcal{M}_{\sigma}$.

If $\underline{\mathit{1}}$ is admissible, then it is the optimal policy and Theorem~\ref{struct} holds with $N_1{=}T{+}1$,
$N_0{=}0$ and $\kappa_{N_1}{=}1$.

Assume now that $\underline{\mathit{1}}$ is not admissible, \emph{i.e.}, $\Delta(\underline{\mathit{0}},\underline{\mathit{1}}){>}\sigma$.
If the optimization problem is feasible, then there exists a policy $\underline{\kappa}^{(1)}{\in}\mathcal{M}_r{\cap}\mathcal{M}_{\sigma}$. 
Starting from $\underline{\kappa}^{(1)}$, it is possible to construct a sequence of policies $\underline{\kappa}^{(1)},\underline{\kappa}^{(2)},
\ldots$ in $\mathcal{M}_r{\cap}\mathcal{M}_{\sigma}$ such that $\mathcal{W}_{\rm S}(\underline{\kappa}^{(k+1)}){>}\mathcal{W}_{\rm S}(\underline{\kappa}^{(k)})$ and converging to the optimal policy $\underline{\kappa}^{(K)}$, where $\underline{\kappa}^{(k+1)}$ has the
structure described in Theorem~\ref{struct}.

Consider a policy $\underline{\kappa}^{(k)}\in\mathcal{M}_r{\cap}\mathcal{M}_{\sigma}$ and
fix $r=\max\{\theta {:} \kappa^{(k)}_{\theta} > 0\}$, \emph{i.e.}, $r$ is the largest state with a non-zero transmission probability.

\begin{figure*}[t]
	\centering
	\includegraphics[width=1.4\columnwidth]{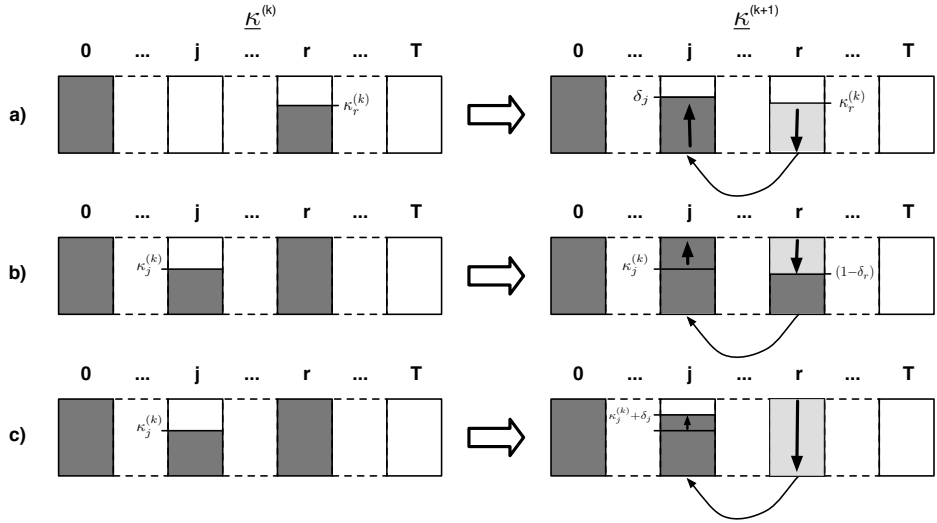}
\caption{Policies $\underline{\kappa}^{(k)}$ and $\underline{\kappa}^{(k+1)}$.}
\vup
\label{structfig}
\end{figure*}

Assume $\kappa^{(k)}_r{<}1$, \emph{i.e.}, randomization occurs in state $r$. Then, the policy in any state $\theta{\neq}r$ is either
$\kappa^{(k)}_{\theta}{=}1$ or $\kappa^{(k)}_{\theta}{=}0$.
If $\exists \theta {:} \theta{<}r{,} \kappa_{\theta}^{(k)} {=} 0$, \emph{i.e.}, the transmission probability in $\theta$ is zero,
then define $\underline{\kappa}^{(k+1)}{=}\underline{\kappa}^{(k)}{+}\underline{u}_j \delta_j {-}\underline{u}_r \kappa^{(k)}_r$
(Fig.~\ref{structfig}.a),
where $j{=}\min \{\theta:\kappa_{\theta}^{(k)} {=} 0\}$ and
with $\delta_j{>}0$ such that
\begin{equation}
\mathcal{J}_{\rm P}(\underline{\kappa}^{(k)}{+}\underline{u}_j \delta_j {-}\underline{u}_r \kappa^{(k)}_r)=\mathcal{J}_{\rm P}(\underline{\kappa}^{(k)}).\end{equation}
We observe that such a $\delta_j$ always exists, due to the continuity of the cost function, Theorem~\ref{th1} and the fact that
$\partial \mathcal{J}_{\rm P}(\underline{\kappa})/\partial \kappa_j>\partial \mathcal{J}_{\rm P}(\underline{\kappa})/\partial \kappa_r$.\footnote{This
intuitive inequality, not proved herein, can be derived by using the expression for the partial derivatives reported in Appendix~\ref{app1}.}
Note that $\underline{\kappa}^{(k+1)}{\in}\mathcal{M}_r{\cap}\mathcal{M}_{\sigma}$. Moreover,
$\mathcal{W}_{\rm S}(\underline{\kappa}^{(k+1)}){>}\mathcal{W}_{\rm S}(\underline{\kappa}^{(k)})$. In fact,
define the policy $\underline{\kappa}^*{=}\underline{\kappa}^{(k)}{-}\underline{u}_r \kappa^{(k)}_r$. Thus,
$\underline{\kappa}^{(k)}{=}\underline{\kappa}^{*}{+}\underline{u}_r \kappa^{(k)}_r$
and $\underline{\kappa}^{(k+1)}{=}\underline{\kappa}^{*}{+}\underline{u}_j \delta_j$ (Fig.~\ref{structfig}.b). Since $\kappa^{*}_j{=}\kappa^{*}_r{=}0$, $\kappa^*_{\theta}{=}0$,
$\forall \theta{>}r$ and $\mathcal{J}_{\rm P}(\underline{\kappa}^{(k)}){=}\mathcal{J}_{\rm P}(\underline{\kappa}^{(k+1)})$, then,
according to Theorem~\ref{mainth},
$\mathcal{W}_{\rm S}(\underline{\kappa}^{(k+1)}){>}\mathcal{W}_{\rm S}(\underline{\kappa}^{(k)})$.
Note that $\underline{\kappa}^{(k+1)}$ is obtained from $\underline{\kappa}^{(k)}$ by \emph{draining} transmission
probability in state $r$, and \emph{pumping} it into state $j{<}r$. In fact, $\kappa^{(k+1)}_j{=}\delta_j{>}\kappa^{(k)}_j{=}0$
and $\kappa^{(k)}_r{>}\kappa^{(k+1)}_r{=}0$.

If $\kappa^{(k)}_r{=}1$ then there may exist a state $\theta{<}r$ such that $0{<}\kappa^{(k)}_{\theta}{<}1$, \emph{i.e.}, the randomization occurs
in state $\theta$. If such a state does not exist, \emph{i.e.}, the map in all states $\theta{<}r$ is deterministic, and there
exists instead at least one state $\theta {:} \theta{<}r{,} \kappa_{\theta}^{(k)} {=} 0$, then fix $j{=}\min \{\theta:\kappa_{\theta}^{(k)} {=} 0\}$.
$\underline{\kappa}^{(k+1)}$ is then constructed from $\underline{\kappa}^{(k)}$ as described before, and via the same considerations
it can be shown that it achieves an improved throughput.

If $\kappa^{(k)}_r{=}1$ and $\exists \theta{<}r : 0{<}\kappa^{(k)}_{\theta}{<}1$, then we fix $j{=}\theta$.
If there exists $0{<}\delta_r{\leq}\kappa^{(k)}_r{=}1$ such that
\begin{equation}
\mathcal{J}_{\rm P}(\underline{\kappa}^{(k)}+\underline{u}_j (1{-}\kappa^{(k)}_j){-}\underline{u}_r \delta_r)=\mathcal{J}_{\rm P}(\underline{\kappa}^{(k)}),
\end{equation} 
\emph{i.e.}, there exists a policy obtained by decreasing the transmission probability in $r$ and setting the transmission probability in
$j$ to unity which incurs the same average cost of policy $\underline{\kappa}^{(k)}$, then, $\underline{\kappa}^{(k+1)}$
is defined as $\underline{\kappa}^{(k+1)}{=}\underline{\kappa}^{(k)}+\underline{u}_j (1{-}\kappa^{(k)}_j){-}\underline{u}_r \delta_r$ (see Fig.~\ref{structfig}.b). We then define
the policy $\underline{\kappa}^{*}{=}\underline{\kappa}^{(k)}{+}\underline{u}_j(1{-}\kappa^{(k)}_j)$.
Thus, since $\kappa^*_j{=}\kappa^*_r{=}1$, $\underline{\kappa}^{(k)}{=}\underline{\kappa}^*{-}\underline{u}_j (1{-}\kappa^{(k)}_j)$,
$\underline{\kappa}^{(k+1)}{=}\underline{\kappa}^*{-}\underline{u}_r \delta_r$, $\kappa^{*}_{\theta}{=}0$, $\forall \theta{>}r$ and
\begin{equation}
\mathcal{J}_{\rm P}(\underline{\kappa}^{*}-\underline{u}_j (1{-}\kappa^{(k)}_j))=\mathcal{J}_{\rm P}(\underline{\kappa}^{*}{-}\underline{u}_r
\delta_r),
\end{equation} 
then, due to Theorem~\ref{mainth2}
\begin{equation}
\mathcal{W}_{\rm S}(\underline{\kappa}^{(k+1)}){>}\mathcal{W}_{\rm S}(\underline{\kappa}^{(k)}).
\end{equation}

Assume now $\nexists \delta_r : 0{<}\delta_r{\leq}\kappa^{(k)}_r{=}1$ such that
\begin{equation}
\mathcal{J}_{\rm P}(\underline{\kappa}^{(k)}+\underline{u}_j (1{-}\kappa^{(k)}_j){-}\underline{u}_r \delta_r)=\mathcal{J}_{\rm P}(\underline{\kappa}^{(k)}),
\end{equation} 
\emph{i.e.}, the cost obtained by setting $\kappa^{(k)}_j{=}1$ and nulling the transmission probability in state $r$ is larger then $\mathcal{J}_{\rm P}(\underline{\kappa}^{(k)})$. In this case, the policy $\underline{\kappa}^{(k+1)}$ is defined as $\underline{\kappa}^{(k+1)}{=}\underline{\kappa}^{(k)}{+}
\underline{u}_j \delta_j{-}\underline{u}_r$ (see Fig.~\ref{structfig}.c), with $0{<}\delta_j{<}1{-}\kappa^{(k)}_j$ and 
\begin{equation}
\mathcal{J}_{\rm P}(\underline{\kappa}^{(k)}+\underline{u}_j \delta_j{-}\underline{u}_r)=\mathcal{J}_{\rm P}(\underline{\kappa}^{(k)}).
\end{equation}
Therefore, the policy $\underline{\kappa}^{(k+1)}$ is obtained from $\underline{\kappa}^{(k)}$ by setting to zero the $r$--th element
of the vector and increasing accordingly the $j$--th element. We show in the following that the reward achieved by policy $\underline{\kappa}^{(k+1)}$
is larger than that achieved by policy $\underline{\kappa}^{(k)}$ also in this case.
The general problem of optimizing $\kappa_j$ and $\kappa_r$ given the transmission probabilities in all the other states
(set according to $\underline{\kappa}^{(k)}$) can be seen as a reduced version of the linear program~(\ref{LP}). The optimal
solution of this reduced problem is 
again a randomized policy with randomization in at most one state, \emph{i.e.}, at least one between $\kappa_j$
and $\kappa_r$ is set to either unity or zero. In the case we are considering, the only two \emph{reduced policies},
corresponding to pairs $(\kappa_j,\kappa_r)$,
with at most one randomization activating the constraint on the maximum performance loss are
$(\kappa^{(k)}_j,\kappa^{(k)}_r)$ and $(\kappa^{(k+1)}_j,\kappa^{(k+1)}_r)$ as defined before. The solution of the reduced LP
is then either $(\kappa^{(k)}_j,\kappa^{(k)}_r)$ or $(\kappa^{(k+1)}_j,\kappa^{(k+1)}_r)$. Fortunately, Theorem~\ref{mainth}
ensures that there exists at least one policy achieving a reward larger than $\underline{\kappa}^{(k)}$ with the
same cost. Therefore, $\underline{\kappa}^{(k)}$ is suboptimal, and $\underline{\kappa}^{(k+1)}$ is optimal. Define
the policy $\underline{\kappa}^*{=}\underline{\kappa}^{(k)}{-}\underline{u}_r(1{-}\kappa^{(k)}_j)$. Thus, $\kappa^*_j{=}\kappa^*_r{=}
\kappa^{(k)}_j$. It can be shown that there exists $\delta^{\prime}_j$, with $0{<}\delta^{\prime}_j{<}1{-}\kappa^{(k)}_j$ such that
\begin{equation}
\mathcal{J}_{\rm P}(\underline{\kappa}^{\prime}){=}\mathcal{J}_{\rm P}(\underline{\kappa}^{(k)}),
\end{equation}
where $\underline{\kappa}^{\prime}{=}\underline{\kappa}^{*}{+}\underline{u}_j \delta^{\prime}_j$. Since $\kappa^*_j{=}\kappa^*_r$,
$\kappa^*_{\theta}{=}0$, $\forall \theta{>}r$, and the above equalities, according to Theorem~\ref{mainth} we have
\begin{equation}
\mathcal{W}_{\rm S}(\underline{\kappa}^{\prime}){>}\mathcal{W}_{\rm S}(\underline{\kappa}^{(k)}).
\end{equation}
As a consequence, $(\kappa^{(k+1)}_j,\kappa^{(k+1)}_r)$ is the optimal solution of the reduced LP introduced above,
and policy $\underline{\kappa}^{(k+1)}$ achieves the maximum reward given the constraint and once fixed the other transmission probabilities.

In all the cases presented, the transmission probability is \emph{drained} from state $r$ and \emph{pumped} into
state $j{<}r$. Note that it is possible to continue the iterations as long as there exists a pair $(j,r): j{<}r, \kappa^{(k)}_j{<}\kappa^{(k)}_r$.
If such indices cannot be found, the iterations terminate with the policy $\underline{\kappa}^{(K)}{\in}\mathcal{M}_r{\cap}\mathcal{M}_{\sigma}$.
It can be easily seen that the iterations terminate with the unique policy in $\mathcal{M}_r{\cap}\mathcal{M}_{\sigma}$ characterized by the structure
indicated in Theorem~\ref{struct}, \emph{i.e.},
\begin{equation}
\underline{\kappa}^{(K)}{=}[\underline{\mathit{1}}_{N_1},~ \kappa_{N_1},~ \underline{\mathit{0}}_{N_0}],
\end{equation}
where $\underline{\mathit{1}}_{N_1}$ and $\underline{\mathit{0}}_{N_0}$ are vectors of $N_1$ ones and $N_0$ zeros,
respectively, and $0{\leq}\kappa_{N_1}{\leq}1$.

Since from any policy $\underline{\kappa}^{(1)}{\in}\mathcal{M}_r{\cap}\mathcal{M}_{\sigma}$ the iterations produce
a policy $\underline{\kappa}^{(K)}{\in}\mathcal{M}_r{\cap}\mathcal{M}_{\sigma}$ such that $\mathcal{W}_{\rm S}(\underline{\kappa}^{(K)}){\geq}
\mathcal{W}_{\rm S}(\underline{\kappa}^{(1)})$, then $\underline{\kappa}^{(K)}$ is the optimal policy, \emph{i.e.},
$\underline{\kappa}^{(K)}{=}\widehat{\underline{\kappa}}$.

Theorem~\ref{struct}, besides unveiling an important feature of the optimal interference control strategy in retransmission-based
systems, also has an immediate practical meaning. In fact, the optimal policy can be computed
through a simple algorithm that generates a sequence $\underline{\kappa}^{(1)},\underline{\kappa}^{(2)},\ldots$ of at most $T$
policies terminating with $\widehat{\underline{\kappa}}$.

Let us fix $\underline{\kappa}^{(1)}=\mathit{\underline{1}}$. If $\Delta(\underline{\mathit{0}},\underline{\mathit{1}}){\leq}\sigma$, then
the optimal policy is $\widehat{\underline{\kappa}}=\underline{\mathit{1}}$. Otherwise, if $\Delta(\underline{\mathit{0}},\underline{\kappa}^{(1)}-\underline{u}_T)\leq\sigma$ the algorithm terminates with the optimal policy $\widehat{\underline{\kappa}}=\underline{\kappa}^{(1)}-\underline{u}_T\delta_T$, where $\delta_T$ is the unique solution of $\Delta(\underline{\mathit{0}},\underline{\kappa}^{(1)}-\underline{u}_T\delta)=\sigma$. If instead $\Delta(\underline{\mathit{0}},\underline{\kappa}^{(1)}-\underline{u}_T)>\sigma$, the algorithm sets $\underline{\kappa}^{(2)}=\underline{\kappa}^{(1)}-\underline{u}_T$, and continues with the next iteration.

Similarly to the previous step, if $\Delta(\underline{\mathit{0}},\underline{\kappa}^{(2)}-\underline{u}_{T{-}1}){\leq}\sigma$ the algorithm
terminates with the optimal policy $\widehat{\underline{\kappa}}{=}\underline{\kappa}^{(2)}-\underline{u}_{T-1}\delta_{T-1}$, where $\delta_{T-1}$
is the solution of $\Delta(\underline{\mathit{0}},\underline{\kappa}^{(2)}-\underline{u}_{T-1}\delta){=}\sigma$. Otherwise, the algorithm sets
$\underline{\kappa}^{(3)}{=}\underline{\kappa}^{(2)}-\underline{u}_{T-1}$ and so on.


Thus, the algorithm sequentially evaluates the variables $\kappa_j$ in decreasing order from $T$ and terminates with the optimal
policy as soon as it finds the first non-zero element.

\begin{figure}[t]
	\centering
	\includegraphics[width=\figw]{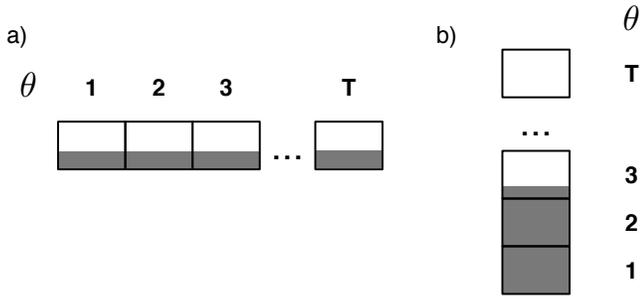}
\caption{Graphical representation of a) the policy in which the secondary source accesses slots in which the primary source transmits
with fixed probability, and b) the optimal policy.}
\vup
\label{flood}
\end{figure}

The structure of the optimal policy leads to another important observation. Consider a secondary source adopting a sensing approach,
such that it always transmits when the channel is sensed idle, and transmits with fixed probability $\kappa$ when the channel is sensed busy 
(see Fig.~\ref{flood}.a). Thus, the secondary source transmits with probability equal to $\kappa$ in all the states ${\theta}$, with $0{<}{\theta}{\leq}T$.
We call this strategy \emph{horizontal flooding}, meaning that the secondary source equalizes the transmission probability such that
it reaches the same level in all the states in which the primary source transmits.

The optimal transmission strategy defined above determines the transmission probabilities in the various states of the
Markov chain under the constraint on the maximum throughput loss of the primary user. As the constraint becomes tighter, the water level of the secondary user is \emph{drained} from the upper states in Fig.~\ref{flood}.b, corresponding to later retransmissions of primary user's packets
in order to reduce the impact of the activity of the secondary user on the primary user's throughput. Note that if $\epsilon{=}0$, the secondary user
is always silent unless the primary user is idle. The optimal policy corresponds
to a \emph{vertical flooding}, where states with a smaller index are flooded with water, \emph{i.e.}, transmission probability, first (see Fig.~\ref{flood}.b).
The horizontal approach, while sometimes simpler to implement, is suboptimal due to Theorem~\ref{mainth}.

Finally, we observe that the arrival rate at the primary source influences the \emph{aggressiveness} of the secondary source. Clearly, as $\alpha$ decreases, also the average throughput of the primary source decreases, as the fraction of time spent sending packets decreases. As the elements of $\underline{\kappa}$ get larger, if $\alpha$ is small, the impact of the increased transmission probability is small, as the secondary source is increasing its access rate in states with low probability. Interestingly, the \emph{fraction} of throughput lost by the primary source decreases as the arrival rate $\alpha$ gets smaller. 

\subsection{Constraint on the Average Failure Probability}
\label{avfprob}
As shown in the previous Section, if the constraint on the average performance loss at the primary transmitter
is defined for the throughput loss, then the optimal policy concentrates transmission and interference of the
secondary source in the first transmissions of the primary source's packets.

Remarkably, the same structure applies to an analogous optimization problem in which the constraint is defined
for the increase of the failure probability of the primary source's packets. 

The average cost is trivially the probability that all the transmissions of a packet fail, \emph{i.e.},
\begin{equation}
\mathcal{J}^{\rm fp}_{\rm P}(\underline{\kappa})=\prod_{t=1}^{T}(\rho+(1-\rho)\lambda\kappa_t).
\end{equation}
The above expression for $\mathcal{J}^{\rm fp}_{\rm P}(\underline{\kappa})$ is also obtained by assigning
the following average cost to the various states: 
\begin{align}
\tilde{\gamma}_{\rm P}(\underline{\kappa},{\theta})&=0~~\forall {\theta}=0,1,\ldots,T{-}1\\
\tilde{\gamma}_{\rm P}(\underline{\kappa},T)&=(\rho+(1-\rho)\lambda\kappa_T)/\pi_{\underline{\kappa}}(1).
\end{align}
In fact, recalling the steady-state probabilities provided in Eq.~(\ref{stdistr}), the resulting average cost is
\begin{eqnarray}
\mathcal{J}^{\rm fp}_{\rm P}(\underline{\kappa})\!\!\!&\!\!{=}\!\!&\!\!\!\sum_{\theta=0}^{T}\!\pi_{\underline{\kappa}}(\theta)\tilde{\gamma}_{\rm P}(\underline{\kappa},{\theta}) {=}\frac{\pi_{\underline{\kappa}}(T)}{\pi_{\underline{\kappa}}(1)}(\rho{+}(1{-}\rho)\lambda\kappa_T)\\
\!\!\!&\!\!{=}\!\!&\!\!\! \prod_{t=1}^{T}(\rho+(1-\rho)\lambda\kappa_t).
\end{eqnarray}
Intuitively, the failure probability is the ratio between the fraction of slots in which the process is in state $T$ \emph{and} a packet fails,
\emph{i.e.}, $\pi_{\underline{\kappa}}(T)\rho_T$,\footnote{We recall that $\rho_T{=}(\rho+(1-\rho)\lambda\kappa_T)$} and the fraction of slots in which the process starts the transmissions of a new packet,
\emph{i.e.}, $\pi_{\underline{\kappa}}(1)$. While the average throughput can be expressed as time average of a sampling function (see
Eq.~(\ref{timeav})), the packet failure probability is then the ratio of the time averages of sampling functions associated with state $F$ and state $1$
multiplied by the failure probability in state $F$.

The cost in state $T$ is, thus, a function of the steady-state distribution. The optimization problem can be reduced to a Linear Program
also in this case. In fact, the constraint on the packet failure probability 
\begin{equation}
\frac{\pi_{\underline{\kappa}}(F)}{\pi_{\underline{\kappa}}(1)}\rho_{T}{\leq}\sigma,
\end{equation}
can be rewritten as $\pi_{\underline{\kappa}}(F)\rho_{T}{-} \pi_{\underline{\kappa}}(1)\sigma{\leq}0$.

Note that the structure of the cost function is significantly different with respect to the throughput case. In fact, under the hypothesis of Theorem~\ref{mainth}, while $\mathcal{J}_{\rm P}(\underline{\kappa}{+}\underline{u}_j\delta){>} \mathcal{J}_{\rm P}(\underline{\kappa}{+}\underline{u}_r\delta)$, in this case the equality holds, \emph{i.e.}, $\mathcal{J}^{\rm fp}_{\rm P}(\underline{\kappa}{+}\underline{u}_j\delta){=}\mathcal{J}^{\rm fp}_{\rm P}(\underline{\kappa}{+}\underline{u}_r\delta)$. Therefore, the overall cost is insensitive to the state in which the secondary source increases the transmission probability. More formally, fix $j$, $r$ and $\delta$, with $0{<}j{<}r{\leq} T$ and $-\min(\kappa_j,\kappa_r){\leq}\delta{\leq}\min(1{-}\kappa_j,1{-}\kappa_r)$, then,
\begin{equation}
\mathcal{J}^{\rm fp}_{\rm P}(\underline{\kappa}+\underline{u}_j\delta)=\mathcal{J}^{\rm fp}_{\rm P}(\underline{\kappa}+\underline{u}_r\delta).
\end{equation}
Interestingly, while the cost in terms of failure probability is insensitive to the state in which the secondary source increases/decreases
the transmission probability, the secondary source's throughput increases faster if the transmission probability is increased in the states
with a small index. These considerations result in an overall behavior of the reward/cost tradeoff analogous to that resulting
from a definition of the primary source's cost in terms of achieved throughput. Then,
Theorems~\ref{mainth} and~\ref{mainth2} hold for this definition of the cost. A detailed proof can be found in Appendix~\ref{appmf}.

Note that again transmission by the secondary source in state $0$ does not have any effect on the cost
of the primary source, while the reward of the secondary source increases as $\kappa_0$ is increased, thus the optimal
value for $\kappa_0$ is one.

 As $\mathcal{J}_{\rm P}(\underline{\kappa})$, also the average cost $\mathcal{J}^{\rm fp}_{\rm P}(\underline{\kappa})$ is a strictly increasing function of any variable $\kappa_{\theta}$, with $\theta{>}0$.
Therefore, for this constraint also, the optimal policy lies in the set $\{\underline{\kappa}{:}\Delta(\underline{\mathit{0}},\underline{\kappa}){=}\sigma\}{\cup}\{\underline{\mathit{1}}\}$.

Since Theorems~\ref{mainth} and~\ref{mainth2} hold, then it is possible to construct a sequence 
of randomized policies achieving an improved reward and converging to the optimal randomized policy
defined in Theorem~\ref{struct}. Therefore, the optimal policy has the same structure of the optimal policy found
for the previously considered case.

Other constraints, as well as other secondary source's performance metrics, may lead to a different optimal policy structure. For
instance, the activity of the secondary source may be limited by a constraint on the average number
of transmissions\footnote{This performance metric is sometimes referred to as delay in the technical literature.} of the packets of the
primary source. For this metric, the average cost of the primary source is
\begin{equation}
\mathcal{J}_{\rm P}^{ntx}(\underline{\kappa})=1{+}\sum_{t=1}^{T-1}\prod_{i=1}^t (\rho+(1-\rho)\lambda\kappa_i).
\end{equation}
Observe that an increased transmission probability in state $j$ increases all the terms of the sum with $t{\geq}j$. 

Similarly to the throughput case, the cost increase associated to an increased transmission probability of the secondary source in state $j$ is
larger than the same increase in state $r{>}j$. However, the difference between the average costs associated with the resulting policies may be larger than in the throughput case. Therefore, for some regions of the parameters, the secondary source may be forced to concentrate its transmissions
in the last transmissions of the primary source's packets.

The optimization problem admits an analogous formulation, and it is possible to derive the structure of the optimal policy by following
a logical procedure entirely similar to the one presented before.

\section{Discussion for the General Case}
\label{gencase}

The structure shown before holds if $\nu^*{=}\nu$, \emph{i.e.}, if primary source's transmission does not alter the decoding probability
at the secondary receiver. The reward collected by the secondary source associated with transmission in state $0$ or state $\theta{>}0$
is then the same. This assumption may fit some configurations of the network and receiver capabilities, \emph{e.g.}, the secondary source is
much closer to the secondary receiver than the primary source, or the secondary receiver can effectively decode and cancel the signal from the
primary source.

However, in general, $\nu^*{\geq}\nu$. In the following the case $\nu^*{>}\nu$ is discussed. This means that if $\kappa_0{=}\kappa_{t}$,
${t}{>}0$, the average reward of the secondary source
in state $0$ is larger than the reward in $t$. In fact, recalling that $\tilde{\omega}_S(\kappa_{\theta},{\theta})$ is the average reward collected by the secondary source in $\theta$ if the transmission probability is $\kappa_{\theta}$, we have:
\begin{equation}
\tilde{\omega}_S(\kappa_0,{0}){=} (1{-}\nu) \kappa_0 {=} (1{-}\nu) \kappa_t{>} (1{-}\nu^*) \kappa_t{=}\tilde{\omega}_S(\kappa_t,{t}). 
\end{equation}
Note that the observations made before on the average cost of the primary source remain valid. The average cost is a monotonic increasing
function of the transmission probabilities and for $0{<}\delta {<} 1{-}\max(\kappa_j,\kappa_r)$ and $0{<}j{<}r{\leq} T$, the following holds:
\begin{equation}
\mathcal{J}_{\rm P}(\underline{\kappa}+\underline{u}_j \delta) > \mathcal{J}_{\rm P}(\underline{\kappa}+\underline{u}_r \delta),
\end{equation}
for any $\underline{\kappa}$.

Interference due to primary source's transmission at the secondary receiver makes state $0$ more \emph{desirable} to the 
secondary source. As observed before, interference increases the average number of transmissions of the primary source's packets.
Therefore, the activity of the secondary source reduces the fraction of slots spent by the primary source in the idle state.
\begin{figure*}[!t]
\normalsize
\setcounter{mytempeqncnt}{\value{equation}}
\setcounter{equation}{43}
\begin{equation}
\label{der1}
\frac{\partial \mathcal{W}_{\rm S}(\underline{\kappa})}{\partial \kappa_1}{=}\frac{\alpha (1 - \lambda (1 - 
      \kappa_2 - \alpha - \nu + \alpha \nu) (1 - \rho) + \alpha \rho - \nu^* (1 + 
      (1 - \rho)\lambda\kappa_2   + \alpha \rho))}{(1 + \alpha  (\rho{+}(1 - \rho)\lambda\kappa_1))^2}
\end{equation}
\begin{equation}
\label{thnu}
\nu^*<\frac{1 - \lambda(-1 + \kappa_2 + \alpha + \nu - \alpha \nu) (-1 + \rho) + \alpha \rho}{1 +  (1 - \rho)\lambda \kappa_2  + \alpha \rho}
\end{equation}
\vspace*{1pt}
\hrulefill
\setcounter{equation}{\value{mytempeqncnt}}
\end{figure*}

Depending on $\nu^*$, $\nu$, $\rho^*$ and $\rho$, an increased transmission in a state $\theta{>}0$ may decrease the average
throughput of the secondary source. Some insights can be extrapolated through the analysis of the case $T{=}2$, \emph{i.e.},
the primary receiver transmits the packets at most twice.
The average reward of the secondary source is
\begin{equation}
\mathcal{W}_{\rm S}(\underline{\kappa}) {=} \frac{(1{-}\alpha)(1{-}\nu)\kappa_0 {+} (1{-} \nu^*)\alpha(  \kappa_1 {+} (\rho{+}(1{-}\rho)\lambda\kappa_1)\kappa_2)   }{1+\alpha (\rho + (1-\rho)\lambda \kappa_1)}.
\end{equation}

$\mathcal{W}_{\rm S}(\underline{\kappa})$ is a monotonically increasing function of $\kappa_0$, irrespective of $\kappa_1$
and $\kappa_2$. In fact,
\begin{equation}
\frac{\partial \mathcal{W}_{\rm S}(\underline{\kappa})}{\partial \kappa_0}= \frac{(1 - \alpha) (1-\nu) }{1 + \alpha( 
 \rho+ (1 - \rho) \lambda \kappa_1)},
\end{equation}
which is trivially positive for any admissible set of parameters. The secondary source's transmission in state $0$ does not modify
the transition probabilities of the Markov chain. Therefore, any increase of $\kappa_0$ corresponds to an increased average reward,
and since it does not influence the cost, again it is optimal to set $\kappa_0{=}1$.

Similar considerations apply to $\kappa_2$, and, more generally, to transmission in state $T$. We obtain,
\begin{equation}
\frac{\partial \mathcal{W}_{\rm S}(\underline{\kappa})}{\partial \kappa_2}= 1 - \nu^* - \frac{
 1 - \nu^*}{1 +  \alpha (\rho+ (1 - \rho)\lambda \kappa_1)},
\end{equation}
which is positive independently of $\kappa_0$ and $\kappa_1$.

Transmission in state $1$, instead, alters the transition probabilities, and increases the fraction of
time spent by the primary source in state $2$, while reducing the time spent in states $0$ and $1$.
The total time spent in the absence of interference from the primary source, which is,
\begin{equation}
\frac{1{-}\alpha}{1+\alpha(\rho{+}(1-\rho)\lambda\kappa_1)},
\end{equation}
decreases as $\kappa_1$ is increased. If the secondary receiver incurs a high failure probability
when decoding a signal interfered by the primary source, the average reward of the secondary 
source may suffer because of the larger average number of transmissions of the primary source's packets
due to transmission in state $1$. The derivative $\partial \mathcal{W}_{\rm S}(\underline{\kappa})/\partial \kappa_1$
is shown in Eq.~(\ref{der1}) and is positive if $\nu^*$ is smaller than the threshold in Eq.~(\ref{thnu}).\footnote{$\kappa_0$
is set to one in the equations.} \addtocounter{equation}{2}

There are thus regions of the parameters and transmission probability $\kappa_2$, such that an increased transmission
probability in state $1$ results in a smaller average reward. Note that an upper bound for $\partial \mathcal{W}_{\rm S}(\underline{\kappa})/\partial \kappa_1$ is obtained by setting $\kappa_2{=}1$. In fact, transmission in state $1$ increases the steady-state probability of state $2$, while transmission in the latter state does not modify the steady-state distribution.\footnote{In general, an upper bound is obtained by setting $\kappa_T{=}1$.} It is easy to see that the threshold in Eq.~(\ref{thnu}), if computed with $\kappa_2{=}1$, becomes smaller than or equal to $1$
for any admissible set of parameters. Therefore, there exists a region of parameters such that the derivative of the average reward 
with respect to $\kappa_1$ is negative. In this region, any throughput-optimal policy sets $\kappa_1{=}0$. The optimal policy may, therefore,
have a different structure than the one shown before for the case $\nu^*{=}\nu$.
In particular, note that the optimal policy may not belong to the set of policies $\{\underline{\kappa}~:~\Delta(\underline{\mathit{0}},\underline{\kappa}){=}\sigma\}\cup\underline{\mathit{1}}$, \emph{i.e.}, the optimal policy may provide a performance reduction to the primary source smaller than the maximum allowed.

In general. if $\nu^*{>}\nu$ the mutual interaction between the activity of the secondary source and that of the primary source becomes
more involved, and it is hard to provide a structure for the optimal policy. 
Intuitively, the larger $\nu^*$, the smaller the transmission probabilities in states $\theta{=}1,2,\ldots,T$, as the secondary source may 
maximize its own throughput by preserving the steady-state probability of state $0$. The same reason may force the secondary source to
concentrate its transmissions in the states corresponding to the last transmissions of a primary source's packet. 

Numerical results illustrating the above discussion are shown in the Section~\ref{numres}.
\section{Online Approaches: State Observation and Model Knowledge}
\label{sec:online}

The resolution of the linear program of Eq.~(\ref{LP}) necessitates the knowledge of the transition probability kernel as well of the cost
functions. However, it can be observed that a relatively small number of parameters (the failure probabilities $\rho$, $\rho^*$, $\nu$ and $\nu^*$, and the arrival probability $\alpha$) determine the transition probability matrix and the cost functions. Therefore,
the estimation of the statistics of the stochastic process and of the cost functions is faster than in a totally unstructured environment.

The realization of the policy requires the perfect identification of the state of the primary user. In the network considered
herein, the estimation of the state within the state space can be obtained by combined channel sensing and packet header decoding.
In fact, the secondary user can distinguish state $0$ from any other transmission state $\Theta{>}0$ by sensing the channel and
detecting the presence of a signal. The header of the packets transmitted by the primary user contains their sequence number.
Therefore, by decoding the header the secondary user can count retransmissions of the same packet. 

By decoding packet header and ACK/NACK feedback sent by the primary and secondary receivers, the secondary user can estimate the transition
probability matrix and the cost functions, as well as identify the state of the primary user. Note that, as the decoding of packet headers and ACK/NACK is crucial to establish communications and instrumental for distributed access mechanisms, these packets are generally
strongly encoded and available to all the neighbors of a node.

If the statistics of the Markov chain and the cost functions are unknown, under the assumption of idealized state observation, reinforcement
learning algorithms~\cite{reinf1} can be employed to iteratively converge to the optimal strategy based on a sample path of observations.
The convergence rate of learning algorithms decreases as the state space gets larger. However, techniques which approximate the learned
functions may speed up the learning rate~\cite{mihaela1}.

In more complex network scenarios the exact identification of the state of the network, as well as the estimation of the statistics of the stochastic process which models its temporal evolution, might be very challenging. 
As observed in some recent work which extends the framework presented herein to online learning~\cite{myglob1,myglob2}, the secondary user
may get access to only some features of the state space. For instance, if the primary user stores
packets in a buffer the number of packets in the buffer is hidden to the secondary user. If the secondary user fails to decode the
header of a primary user's packet it may detect the presence of a signal but the retransmission index remains unknown. Another example
of hidden state variable is the channel state of the primary links. Channel knowledge would increase the effectiveness of secondary user
transmission. In fact, the secondary user can potentially reduce the impact of the generated interference by scheduling transmissions in those time
slots in which the link between the primary transmitter and the primary receiver is very strong, and thus interference would not impair packet
reception, or very weak, and thus the primary user packet would fail in any case. In the absence of channel state information, the secondary
user bases its decision making on the average effect of actions over channel states, that is, the failure probability associated with idleness and
transmission. Analogously, if a backoff mechanism is implemented by the primary users to regulate channel access the secondary user may be
unable to distinguish between idleness due to empty buffer or backoff. 

In general, by observing the operations of the nodes it is possible to acquire a significant amount of information about the state of the network.
The amount of information collectible by the secondary user depends on the transmission, access and networking protocols. For instance, 
the rigid access structure provided by Time Division Multiple Access (TDMA) provides more information to the observer than random access. In fact,
an idle TDMA slot means that the assigned user has an empty buffer, whereas idleness in random access may be related to the access mechanism
itself.

If the statistics of the process and the state-observation map are known to the secondary user,
then the secondary user can base its decision on a belief vector~\cite{klc} collecting the maximum likelihood distribution of the real state of the system.
Since a priori knowledge of statistics and state-observation map is unrealistic in general scenarios, the approach proposed in~\cite{myglob2} is to optimize the distribution of the states in the observation space based on the estimated cost functions, which collects all the possible observations. 

According to this discussions, the framework presented in this paper opens many exciting new areas of investigation.

\section{Numerical Results}
\label{numres}


In this Section, numerical results validating the findings and observations made throughout the paper are presented.
We recall that $\alpha$ is the probability that the primary source transmits a fresh packet in a slot not allocated to packet retransmission;
$\rho$ is the probability that the primary receiver correctly decodes a packet sent by the primary source in a slot in which the secondary source 
is silent. The failure probability at the primary receiver if the secondary source transmits is $\rho^*{=}\rho{+}\lambda(1{-}\rho)$, where
$\lambda$ is the failure probability increase. $\nu$ and $\nu^*$ are the failure probability at the secondary receiver if the
primary source is silent and transmits, respectively. A failure probability increase is also defined for the secondary receiver
by $\lambda_S$ such that $\nu^*{=}\nu{+}\lambda_S(1{-}\nu)$.

In Sections~\ref{nr_tc} and~\ref{nr_fc}, we present numerical results for $\nu^*{=}\nu$, \emph{i.e.},
the Z-channel, where the constraint is defined on throughput and failure probability, respectively.
Section~\ref{nr_int} presents numerical results for the case $\nu^*{>}\nu$.

In all the following plots, the maximum number of transmissions of a primary source's packet is fixed to $T{=}4$.
 \begin{figure}[t]
	\centering
	\includegraphics[width=.96\columnwidth]{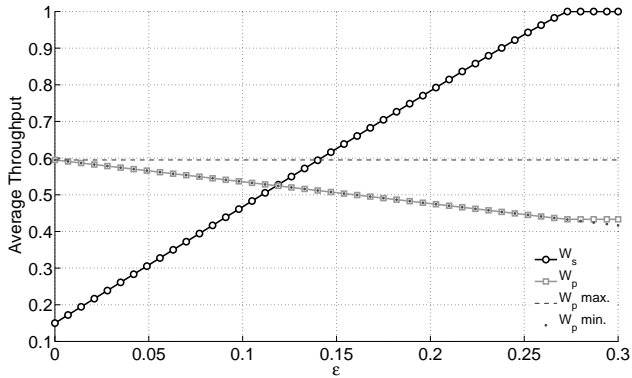}
\caption{Throughput as a function of the maximum fraction of throughput loss, where $\alpha{=}0.8$, $\rho{=}0.3$, $\nu{=}\nu^*{=}0$ and $\lambda{=}0.3$.}
\vup
\label{sig}
\end{figure}
\subsection{Constraint on the primary source's throughput, $\nu^*{=}\nu$}
\label{nr_tc}
In this Section, numerical results for the Z-channel network with a constraint on the throughput loss of the primary source are presented.
The performance loss is parameterized through $\epsilon$, defined as the maximum \emph{fraction} of throughput loss of the primary source,
\emph{i.e.}, the maximum throughput loss is $\sigma{=}\mathcal{W}_{\rm P}(\underline{\mathit{0}})\epsilon{=}(1{-}\mathcal{J}_{\rm P}(\underline{\mathit{0}}))\epsilon$.

\begin{figure}[t]
	\centering
	\includegraphics[width=.96\columnwidth]{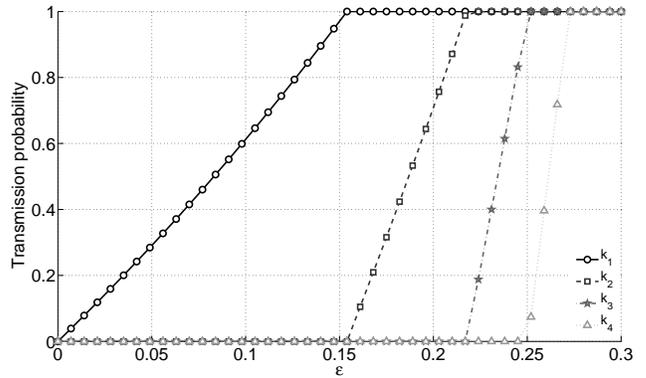}
\caption{Transmission probabilities as a function of the maximum fraction of throughput loss, where $\alpha{=}0.8$, $\rho{=}0.3$, $\nu{=}\nu^*{=}0$ and $\lambda{=}0.3$.}
\vup
\label{sigk}
\end{figure}

In Figs. \ref{sig} and \ref{sigk}, the throughput and the secondary source's transmission probability are depicted as a function of $\epsilon$. In the picture, $\mathcal{W}_{\rm P} {\rm max}$ and $\mathcal{W}_{\rm P} {\rm min}$ correspond to the throughput achieved by the primary source when the secondary source
is silent and the minimum throughput of the primary source according to the constraint.\footnote{In this and in the following Section, the throughput of the secondary source is normalized to $(1{-}\nu)$.}

The throughput of the secondary source increases as $\epsilon$ is increased. A larger $\epsilon$ allows the
secondary source to interfere more with the primary source. The throughput actually achieved by the primary source decreases according
to the increased maximum performance loss allowed, and it can be observed that the policy of the secondary source lowers the throughput
of the primary one as much as possible in order to maximize the secondary throughput. 
When the throughput of the secondary source is equal to one, corresponding to the former transmitting with probability one in every slot,
the throughput of the primary source stops decreasing, as the secondary source cannot interfere more.


Fig. \ref{sigk} shows that the policy
of the secondary source follows the structure discussed before. Thus, with $\epsilon{=}0$
the secondary source is allowed to transmit only in the slots where the primary is not accessing the channel ($\kappa_0{=}1$ and $\kappa_t{=}0$, $t{>}0$).
As $\epsilon$ increases, the transmission probability in state $1$, \emph{i.e.}, $\kappa_1$, increases until it reaches unity. Then,
$\kappa_2$ starts to increase and so on until all the $\kappa_t$'s are set to unity.

The rate increase of the various $\kappa_t$'s is different. In particular, the rate increase of the $\kappa_t$'s corresponding to transmission
in states with small indices is smaller than those corresponding to large indices. In fact, interference in the states
corresponding to the first transmissions of a packet generates a larger primary source's throughput reduction than
interference in the later transmissions. Conversely, the throughput of the primary source gets larger, and so does the maximum
throughput loss.

\begin{figure}[t]
	\centering
	\includegraphics[width=.96\columnwidth]{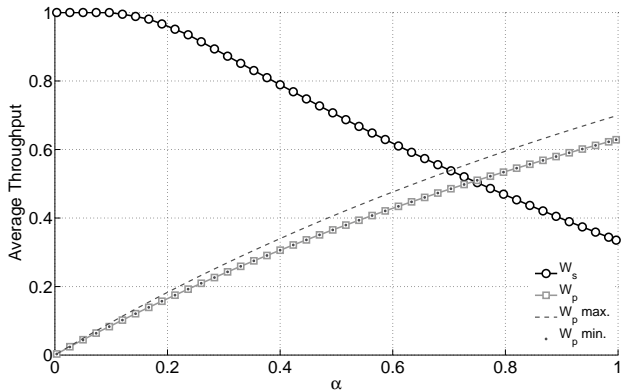}
\caption{Throughput as a function of the arrival rate $\alpha$, where $\epsilon{=}0.1$, $\rho{=}0.3$, $\nu{=}\nu^*{=}0$ and $\lambda{=}0.3$.}
\vup
\label{alp}
\end{figure}

\begin{figure}[t]
	\centering
	\includegraphics[width=.96\columnwidth]{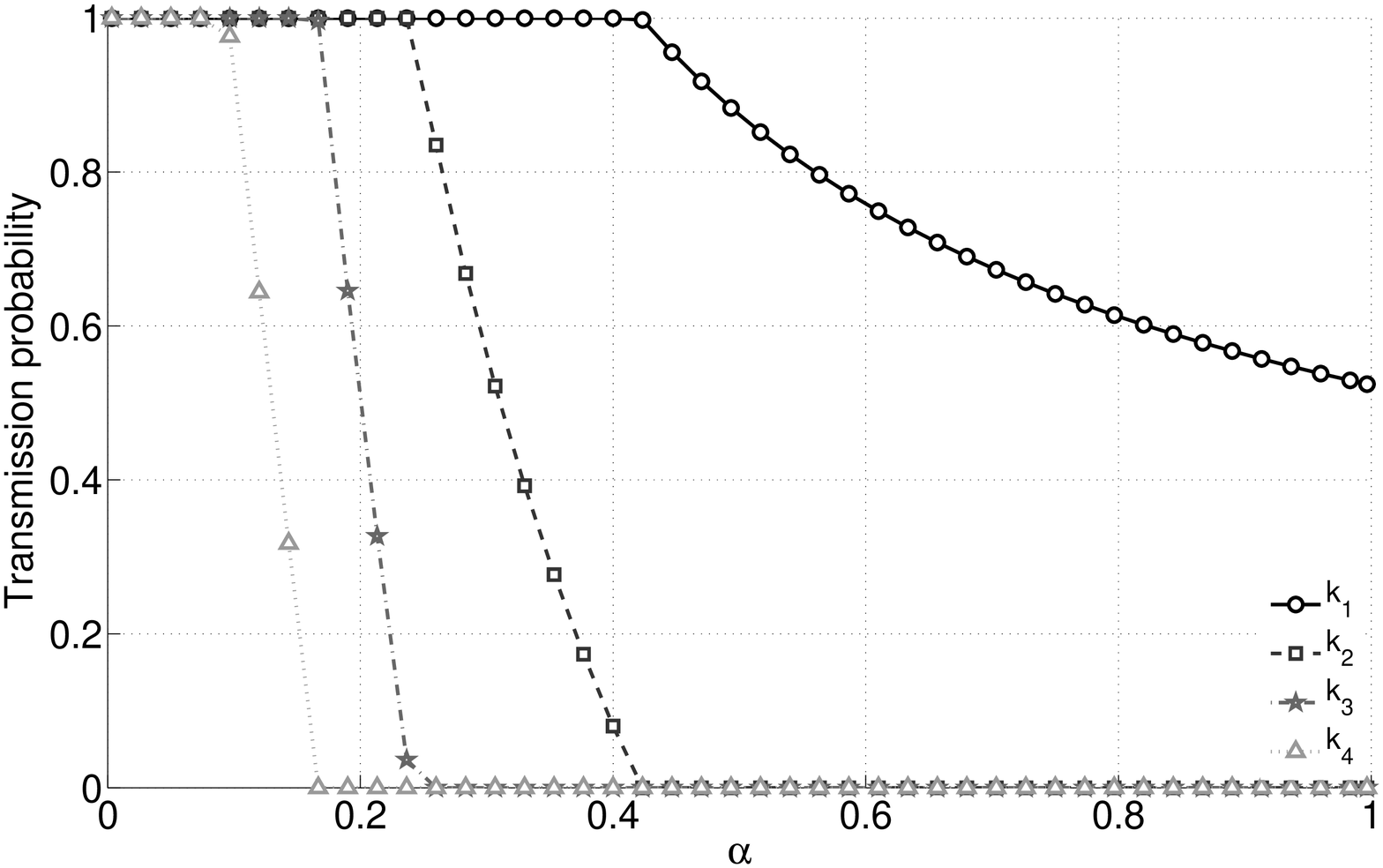}
\caption{Transmission probabilities as a function of the arrival rate $\alpha$, where $\epsilon{=}0.1$, $\rho{=}0.3$, $\nu{=}\nu^*{=}0$ and $\lambda{=}0.3$.}
\vup
\label{alpk}
\end{figure}

\begin{figure}[t]
	\centering
	\includegraphics[width=.96\columnwidth]{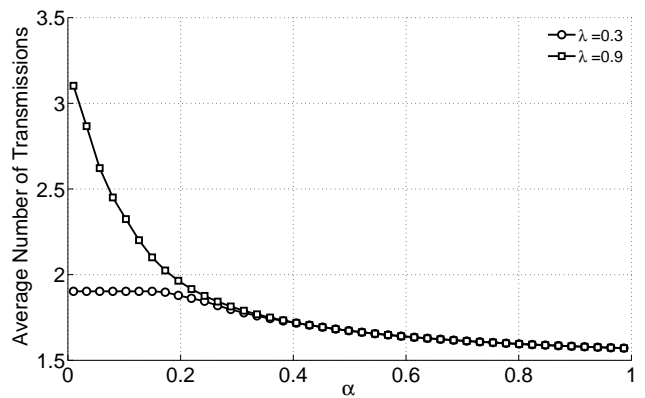}
\caption{Average number of transmissions as a function of $\alpha$, where $\epsilon{=}0.1$, $\rho{=}0.3$, $\nu{=}\nu^*{=}0$, $\lambda{=}0.3$.}
\vup
\label{th_numtx}
\end{figure}

Figs. \ref{alp} and \ref{alpk} show the same quantities as a function of  $\alpha$, \emph{i.e.}, the arrival rate of new packets at the primary source.
As expected, the throughput of the secondary source decreases as $\alpha$ increases. Fig~\ref{th_numtx} depicts the average number of transmissions
of the primary packets for the same parameters, with $\lambda{=}0.3$ and $\lambda{=}0.9$.

A larger $\alpha$ means that the primary source
is accessing the channel more often. Therefore, the number of slots in which the secondary source can transmit while meeting the constraint on the
throughput loss of the primary source decreases.
However, there is another effect of a large $\alpha$ that needs to be considered besides the scarcity of empty slots (in which the secondary
source transmits with probability one). In fact, if the probability that a fresh packet is transmitted in an idle slot by the primary source is
small, an increased average number of transmissions for each packet has a smaller effect on the throughput of the primary source.
The additional retransmissions forced by the interference are likely to substitute for slots in which the primary source
would be idle anyway, and thus, are the slots in which the primary source would incur the highest possible cost. On the other hand, if $\alpha$ is large, additional retransmissions are performed instead of new packet transmissions that collect an average cost smaller than that of an empty slot.

The relation between $\alpha$ and the interference generated by the secondary source to the primary receiver is illustrated in Fig.~\ref{alpk}, 
and Fig.~\ref{th_numtx}. Fig.~\ref{alpk} shows that the throughput trend of Fig.~\ref{alp} does not only correspond to a smaller fraction of empty slots,
but that the policy in the states $\theta{>}0$ is a function of the arrival rate.
The fraction of slots in which the secondary source superposes its activity with that of the primary source, normalized by the fraction of slots
in which the latter source transmits, decreases as $\alpha$ increases. The explanation for this behavior is illustrated above. If $\alpha$
is small, the primary source is often idle, and the retransmissions induced by interference generate a smaller loss in the throughput of the
primary source. In fact, if $\alpha$ is small, the secondary source is allowed to force more retransmissions
(see Fig.~\ref{th_numtx}).
Note that again the policy follows the structure discussed in the previous section, where the
$\kappa_t$'s sequentially turn off as the secondary source is forced to reduce the interference.
 \begin{figure}[t]
	\centering
	\includegraphics[width=.96\columnwidth]{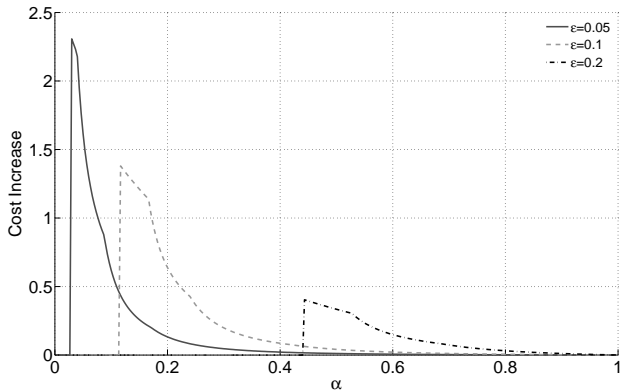}
\caption{Cost increase as a function of $\alpha$, where $\epsilon{=}0.1$, $\rho{=}0.3$, $\nu{=}\nu^*{=}0$, $\lambda{=}0.3$.}
\vup
\label{cinc}
\end{figure}


Finally, Fig.~\ref{cinc} shows $(\mathcal{J}_{\rm S}(\widehat{\underline{\kappa}}_{hf}){-}\mathcal{J}_{\rm S}(\widehat{\underline{\kappa}}))/\mathcal{J}_{\rm S}(\widehat{\underline{\kappa}})$ as a function of the arrival rate $\alpha$, where $\mathcal{J}_{\rm S}(\widehat{\underline{\kappa}}_{hf})$ is the optimal cost for the horizontal flooding approach.\footnote{If $\mathcal{J}_{\rm S}(\widehat{\underline{\kappa}}_{hf}){=}\mathcal{J}_{\rm S}(\widehat{\underline{\kappa}}){=}0$ the ratio is assumed to be equal to zero.} The optimal
transmission probabilities for the horizontal flooding approach are numerically found via a LP slightly more involved than that discussed
herein. Thus, the curves represent the fraction of cost increase when horizontal flooding is adopted instead of vertical flooding. When $\alpha$ is sufficiently small, the cost increase
is zero, as both approaches transmit in all states with probability one. As $\alpha$ increases, the secondary source is forced to reduce the fraction of time in which it transmits in both vertical and horizontal flooding. In the former case, the secondary source starts decreasing the transmission probability in state $T$, while in the latter, the transmission probability is reduced in all states $t{>}0$. However, as soon as $\underline{\mathit{1}}$ becomes
inadmissible, in order to meet the constraint on the 
maximum throughput loss, the horizontal approach is forced to reduce the average transmission time of the secondary source much more quickly
than the vertical approach.
Then, as the arrival rate $\alpha$ is further
increased, the cost increase diminishes, since the advantage due to the concentration of the interference in states with smaller indices vanishes.
In fact, vertical flooding improves the delivery probability of a packet at the expense of a larger average number of transmissions.

\subsection{Constraint on the primary source's failure probability, $\nu^*{=}\nu$}
\label{nr_fc}
In this Section, results for the optimization problem with a constraint defined on the failure probability of the primary source's packets
are presented.\footnote{We remark that by failure probability of a packet, we refer to the probability that all the $T$ transmissions fail.} The activity of the secondary source
increases the failure probability, and the maximum failure probability increase allowed is given by $\sigma$. This increase, $\sigma$, is again parameterized
through $\epsilon$, defined as the maximum relative failure probability increase, that is, $\sigma{=}\mathcal{J}^{\rm fp}_{\rm S}(\underline{\mathit{0}})(1{+}\epsilon)$.

\begin{figure}[t]
	\centering
	\includegraphics[width=.96\columnwidth]{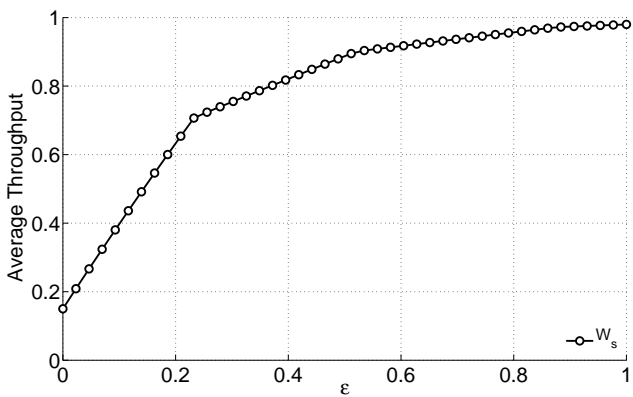}
\caption{Throughput of the secondary source as a function of the maximum performance loss $\epsilon$, where $\alpha{=}0.8$. $\rho{=}0.3$, $\nu{=}\nu^*{=}0$ and $\lambda{=}0.1$.}
\vup
\label{fp_epsth}
\end{figure}

\begin{figure}[t]
	\centering
	\includegraphics[width=.96\columnwidth]{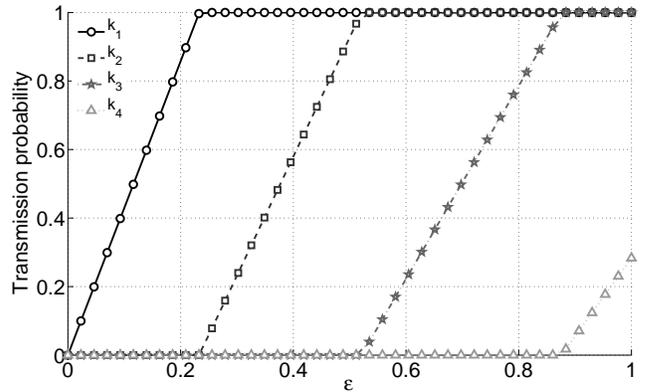}
\caption{Transmission probabilities as a function of the maximum performance loss $\epsilon$, where $\alpha{=}0.8$. $\rho{=}0.3$, $\nu{=}\nu^*{=}0$ and $\lambda{=}0.1$.}
\vup
\label{fp_epsk}
\end{figure}

\begin{figure}[t]
	\centering
	\includegraphics[width=.96\columnwidth]{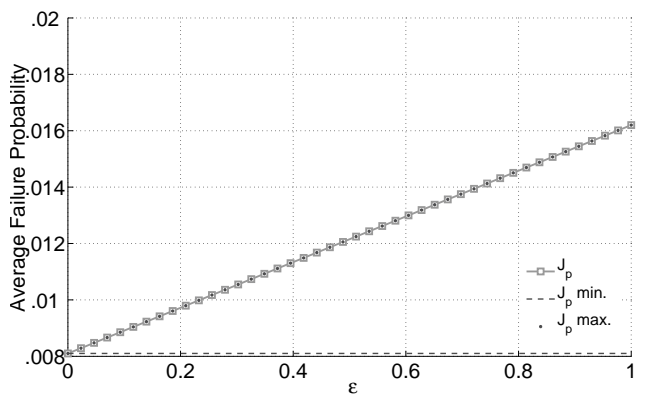}
\caption{Packet failure probability as a function of the maximum performance loss $\epsilon$, where $\alpha{=}0.8$. $\rho{=}0.3$, $\nu{=}\nu^*{=}0$ and $\lambda{=}0.1$.}
\vup
\label{fp_epsfp}
\end{figure}

Figs.~\ref{fp_epsth},~\ref{fp_epsk} and~\ref{fp_epsfp} show the throughput of the secondary source, the secondary source transmission probability, and the failure probability as a function of $\epsilon$, respectively. In Fig.~\ref{fp_epsfp}, $\mathcal{J}_{\rm min}$ is the failure probability associated with 
an idle secondary source, and $\mathcal{J}_{\rm max}$ is the maximum failure probability according to the constraint.
 
Intuitively, the throughput, as well as the overall fraction of slots in which the secondary source transmits, increase as the
maximum failure probability of the primary source's packets increases. The transmission strategy of the secondary source follows the structure
discussed throughout the paper. As the constraint becomes less stringent, first transmission in state $0$ is increased, then transmission
in state $1$ and so on, until the secondary source transmits with probability one in all states.

\begin{figure}[t]
	\centering
	\includegraphics[width=.96\columnwidth]{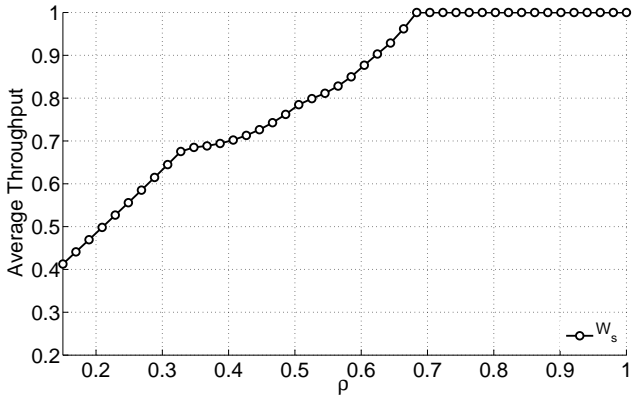}
\caption{Throughput of the secondary source as a function of the failure probability of the primary source in the absence of interference $\rho$, where $\alpha{=}0.8$. $\rho{=}0.3$, $\nu{=}\nu^*{=}0$ and $\lambda{=}0.1$.}
\vup
\label{fp_rhoth}
\end{figure}

\begin{figure}[t]
	\centering
	\includegraphics[width=.96\columnwidth]{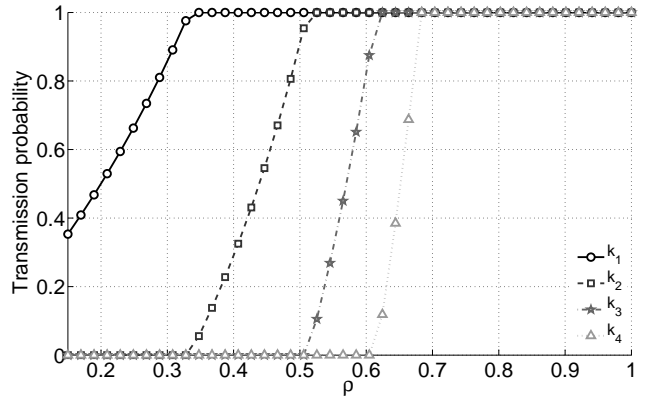}
\caption{Transmission probabilities as a function of the failure probability of the primary source in the absence of interference $\rho$, where $\alpha{=}0.8$. $\rho{=}0.3$, $\nu{=}\nu^*{=}0$ and $\lambda{=}0.1$.}
\vup
\label{fp_rhok}
\end{figure}

\begin{figure}[t]
	\centering
	\includegraphics[width=.96\columnwidth]{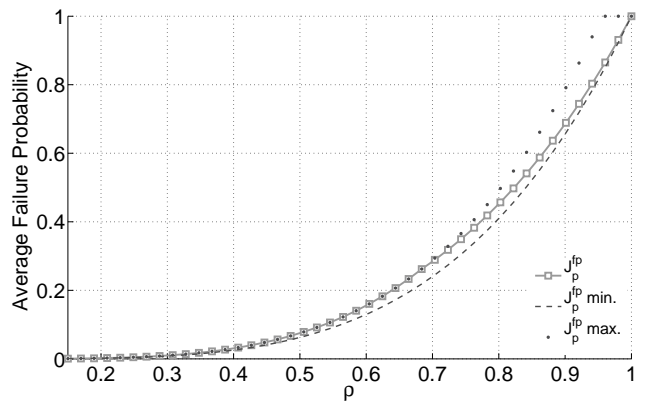}
\caption{Packet failure probability as a function of the failure probability of the primary source in the absence of interference $\rho$, where $\alpha{=}0.8$. $\rho{=}0.3$, $\nu{=}\nu^*{=}0$ and $\lambda{=}0.1$.}
\vup
\label{fp_rhofp}
\end{figure}

Figs.~\ref{fp_rhoth},~\ref{fp_rhok} and~\ref{fp_rhofp} provide the same metrics of the previous figures as a function of the failure probability
of the primary source's transmissions $\rho$. Note that the failure probability of the packets of the primary source if the secondary source
is always idle is $\rho^T$.

The throughput of the secondary source, as well as the transmission probabilities in the states $\theta{>}0$, increase
as $\rho$ becomes larger. We observe the following:
\begin{itemize}
\item the maximum $\sigma{=}\rho^T (1{+}\epsilon)$ polynomially increases with $\rho$. This means that the constraint becomes less stringent
as $\rho$ increases (see Fig.~\ref{fp_rhofp});
\item the primary source transmits in a larger fraction of slots as $\rho$ increases, due to a larger average number of retransmissions. The secondary source has fewer empty slots in which to transmit without interfering with the primary source.
\end{itemize}

If $\rho$ is small, and the secondary source keeps idle, the primary source transmits in a fraction of slots close to $\alpha$.
As $\rho$ gets larger, the primary source increases the fraction of slots in which it transmits, because of the retransmissions.
Nevertheless, the constraint becomes less stringent as $\rho$ increases. In fact, the maximum allowed failure probability is
$\sigma{=}\rho^T(1{+}\epsilon)$. Therefore, as $\rho$ increases the secondary source can increase its activity in states $\theta{>}0$.
The tradeoff between those two effects determines the optimal throughput achieved by the secondary source. For the considered 
set of parameters, the increase of $\sigma$ wins over the decrease of the number of empty slots.

\subsection{Case $\nu^*{>}\nu$}
\label{nr_int}
In this Section, illustrative results for the general case $\nu^*{>}\nu$ are shown. As discussed in Section~\ref{gencase},
in this case, the structure of the optimal policy depends on the parameters. In fact, due to the effect of the interference
by the primary source at the secondary receiver, the secondary source may be forced to be silent in states
$\theta{>}0$ in order not to decrease the steady-state probability of the empty-slot state $0$. In the following, the constraint
is defined on the throughput loss of the primary source.

\begin{figure}[t]
	\centering
	\includegraphics[width=.96\columnwidth]{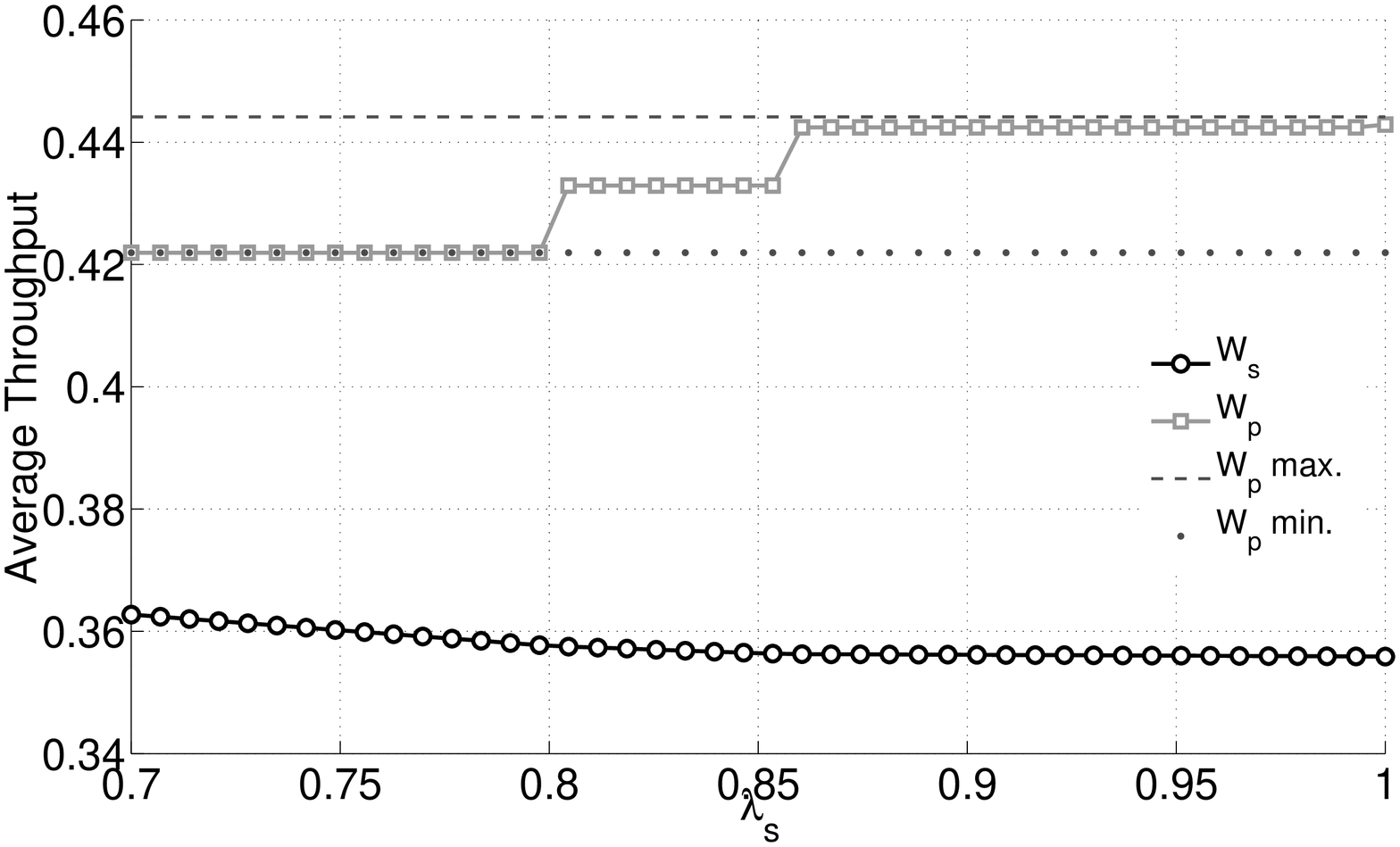}
\caption{Throughput as a function of $\lambda_S, where $ $\alpha{=}0.5$, $\lambda{=}0.6$, $\rho{=}\nu{=}0.2$ and $\epsilon{=}0.05$.}
\vup
\label{int_lamth}
\end{figure}
\begin{figure}[t]
	\centering
	\includegraphics[width=.96\columnwidth]{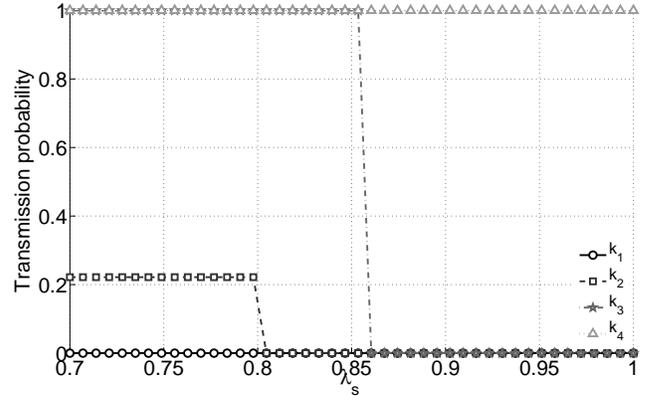}
\caption{Transmission probabilities as a function of $\lambda_S, where $ $\alpha{=}0.5$, $\lambda{=}0.6$, $\rho{=}\nu{=}0.2$ and $\epsilon{=}0.05$.}
\vup
\label{int_lamk}
\end{figure}

The throughput and the transmission probabilities as a function of $\lambda_S$ are depicted in Figs.~\ref{int_lamth}
and~\ref{int_lamk}. We recall that $\lambda_S{\in}[0,1]$ determines how decoding at the secondary receiver is 
hampered by primary source's transmissions. The values $\lambda_S{=}0$ and $\lambda_S{=}1$ correspond to $\nu^*{=}\nu$
and $\nu^*{=}1$, respectively.

As a first observation, the throughput of the secondary source decreases as $\lambda_S$ increases. In fact, the effect of interference
both decreases the reward associated with transmission in the states $\theta{>}0$ and forces the secondary source to reduce its
overall activity. For the same reason, the throughput of the primary source increases and moves close to the maximum throughput,
that is, the secondary source rarely interferes with the primary source.

In fact, as $\nu^*$ gets closer to one, the secondary source reduces transmission, and interference, in states $0{<}\theta{<}T$.
This is done in order to avoid retransmissions, which would reduce the availability of white space.
Interference in state $T$ does not induce a higher probability of further retransmissions. Therefore, $\kappa_T$ remains
one as long as it is admissible according to the constraint. Note that $\kappa_1$, \emph{i.e.}, the transmission probability
in the state which has the largest impact on the average number of retransmissions, is set to $0$. For this configuration
of parameters, the policy takes the opposite form with respect to that described for the case $\nu^*{=}\nu$, \emph{i.e.},
the secondary source concentrates transmissions in the last states of the chain, in order to have a smaller impact on the number of transmissions
of the primary source's packets.
\begin{figure*}[!b]
\normalsize
\vspace*{1pt}
\hrulefill
\setcounter{mytempeqncnt}{\value{equation}}
\setcounter{equation}{50}
\begin{eqnarray}
\label{th1eq}
\frac{\partial \mathcal{N}_{J_{\rm P}} }{\partial \kappa_{\theta}}\mathcal{D}{-}\frac{\partial \mathcal{D}}{\partial 
 \kappa_{\theta}}\mathcal{N}_{J_{\rm P}} \!\!\!&\!\! =\!\!&\!\!\! \bigg(\alpha (1{-}\rho)\lambda \prod_{i=1, i\neq {\theta}}^T (\rho{+}(1{-}\rho)\lambda\kappa_i)\bigg)\bigg(\mathcal{D}+1\bigg){+}\nonumber\\
\!\!\!&\!\! +\!\!&\!\!\! \bigg(1{-}\alpha^2\prod_{i=1}^{T}(\rho{+}(1{-}\rho)\lambda\kappa_i)\bigg)\bigg(\sum_{t={\theta}}^{T-1} (1{-}\rho)\lambda \prod_{i=1, i\neq {\theta}}^{t} (\rho{+}(1{-}\rho)\lambda\kappa_i)\bigg)
%
 %
\end{eqnarray}
\setcounter{equation}{\value{mytempeqncnt}}
\end{figure*}

\section{Conclusions}
\label{concl}
In contrast to much prior work on cognitive networks, in this paper we investigated a scenario wherein the secondary
source is allowed to superpose its transmissions over those of the primary source. The secondary source aims
to maximize its own throughput, while guaranteeing a bounded performance loss for the primary source.
We derived the optimal transmission policy for the secondary user when the primary user adopts
a retransmission based error control scheme. If the decoding probability at the secondary receiver
is not increased by the primary source's transmissions, the resulting optimal strategy of the secondary user has
a unique structure. In particular, the optimal throughput is achieved by the
secondary user by concentrating its interference to the primary user in the first transmissions of
a packet. This is a first step toward a better understanding of interference control strategies in dynamic wireless networks.

\appendices

\section{Proof of Theorem~\ref{th1}}
\label{app1}
\begin{proof}
Theorem~\ref{th1} states that if any component $\kappa_{\theta}$ of the vector $\underline{\kappa}$ is increased, with ${\theta}{>}0$, then $\mathcal{J}_{\rm P}(\underline{\kappa})$ increases. This corresponds to the intuitive fact that a larger transmission probability of the secondary source in any of the states in which the primary source transmits results in a smaller throughput achieved by the latter.

In order to prove this result, we show that $\partial \mathcal{J}_{\rm P}(\underline{\kappa})/\partial \kappa_{\theta}{>}0$, $\forall {\theta}>0$.

Let us introduce the following notation
\begin{eqnarray}
\mathcal{N}_{J_{\rm P}}(\underline{\kappa})\!&\!\!\!{=}\!\!\!&\!(1{-}\alpha){+}\alpha\sum_{t=1}^{T}\prod_{i=1}^{t}(\rho{+} (1 {-} \rho)\lambda \kappa_i)\\
\mathcal{D}(\underline{\kappa})\!&\!\!\!{=}\!\!\!&\!1{+}\alpha\sum_{t=1}^{T-1}\prod_{i=1}^{t}(\rho{+} (1 {-} \rho)\lambda \kappa_i).
\end{eqnarray}
The cost is then $\mathcal{J}_{\rm P}(\underline{\kappa}){=}\mathcal{N}_{J_{\rm P}}(\underline{\kappa})/\mathcal{D}(\underline{\kappa})$.
In the following, the obvious dependence of the above functions on $\underline{\kappa}$ is dropped from the notation.

The derivative of the cost of the primary source can be obtained through the well-known formula
\begin{equation}
\partial \mathcal{J}_{\rm P}/\partial \kappa_{\theta}{=}\frac{   \frac{\partial \mathcal{N}_{J_{\rm P}} }{\partial \kappa_{\theta}}\mathcal{D}{-}\frac{\partial \mathcal{D}}{\partial \kappa_{\theta}}\mathcal{N}_{J_{\rm P}}}{(\mathcal{D})^2}.
\end{equation}

In the previous equation, the denominator is always positive, and thus we focus on the numerator. We have
\begin{eqnarray}
\frac{\partial \mathcal{N}_{J_{\rm P}}}{\partial \kappa_{\theta}} \!&\!\!\!{=}\!\!\!&\! \alpha \sum_{t={\theta}}^{T} (1{-}\rho)\lambda \prod_{i=1, i\neq {\theta}}^t (\rho{+}(1{-}\rho)\lambda\kappa_i) \\
\frac{\partial \mathcal{D}}{\partial \kappa_{\theta}} \!&\!\!\!{=}\!\!\!&\! \alpha \sum_{t={\theta}}^{T-1} (1{-}\rho)\lambda \prod_{i=1, i\neq {\theta}}^{t} (\rho{+}(1{-}\rho)\lambda\kappa_i).
\end{eqnarray}
Through simple algebraic manipulation, we obtain the expression in Eq.~(\ref{th1eq}).\addtocounter{equation}{1}
Since
\begin{equation}
1{-}\alpha^2\prod_{i=1}^{T}(\rho{+}(1{-}\rho)\lambda\kappa_i)>0,
\end{equation}
then all the terms in Eq.~(\ref{th1eq}) are strictly positive.\footnote{Note that in the degenerate cases $\alpha{=}0$, $\lambda{=}0$, or $\rho{=}1$
Eq.~(\ref{th1eq}) is equal to zero.}
Therefore, the derivative is strictly positive and the cost is a monotonically increasing function of any element $\kappa_{\theta}$ with $\theta{>}0$
\end{proof}

\section{Proof of Theorem~\ref{th2}}
\label{app2}
\begin{proof}
Theorem~\ref{th2} states that if the secondary source transmits with a higher probability in any state $\theta$, \emph{i.e.}, $\kappa_{\theta}$ is increased, then the average throughput achieved by the secondary source increases. This may appear a trivial consideration.
However, it must be observed that the transmission probabilities $\kappa_{\theta}$, $0{<}\theta{<}T$ influence the steady-state distribution
of the Markov chain of the network and thus influence the average throughput of the secondary user.

A larger $\kappa_{\theta}$ results in a larger probability that the primary source fails the $\theta$--th transmission of a packet,
and, thus, a larger probability that the Markov process moves to states $\theta{+}1,\ldots,T$. As a consequence, the
steady-state probabilities of the latter states increase, while those of states $0,\ldots,{\theta}$ decrease. Intuition suggests
that, in some cases, a larger $\kappa_{\theta}$ may result in a smaller overall average transmission probability of the secondary
source, \emph{i.e.}, $\sum_t\pi_{\underline{\kappa}}(t)\kappa_t$.
Theorem~\ref{th2} instead ensures that a larger value of any of the $\kappa_{\theta}$ always results in a larger $\sum_t\pi_{\underline{\kappa}}(t)\kappa_t$.  

The proof of this theorem is analogous to that of Theorem~\ref{th1}. In particular, we show in the following that
$\partial \mathcal{W}_{\rm S}(\underline{\kappa})/\partial \kappa_{\theta}{>}0$, $\forall {\theta}{\in}\{0,\ldots,T\}$.

Recalling the definition of $\mathcal{D}(\underline{\kappa})$ given in the previous Theorem, we write $\mathcal{W}_{\rm S}(\underline{\kappa}){=}\mathcal{N}_{W_{\rm S}}(\underline{\kappa})/\mathcal{D}(\underline{\kappa})$,
where
\begin{eqnarray}
\mathcal{N}_{W_{\rm S}}(\underline{\kappa})\!&\!\!\!{=}\!\!\!&\!(1{-}\alpha){+}\alpha\sum_{t=1}^{T}\kappa_{t}\prod_{i=1}^{t-1}(\rho{+} (1 {-} \rho)\lambda \kappa_i).
\end{eqnarray}

In the following, we drop in the notation the dependence between these functions and the policy.
The derivative of the numerator is
\begin{eqnarray}
\frac{\partial \mathcal{N}_{W_{\rm S}}}{\partial \kappa_{\theta}}\!&\!\!\!{=}\!\!\!&\! \alpha\prod_{i=1}^{{\theta}-1}(\rho{+}(1{-}\rho)\lambda\kappa_i){+}\nonumber\\
\!&\!\!\!{+}\!\!\!&\!\alpha\sum_{t={\theta}+1}^{T}(1{-}\rho)\lambda\kappa_t\prod_{i=1,i{\neq}{\theta}}^{t-1}(\rho{+}(1{-}\rho)\lambda\kappa_i).
\end{eqnarray}
Again, the derivative can be written as
\begin{equation}
\partial \mathcal{W}_{\rm S}/\partial \kappa_{\theta}{=}\frac{   \frac{\partial \mathcal{N}_{W_{\rm S}} }{\partial \kappa_{\theta}}\mathcal{D}{-}\frac{\partial \mathcal{D}}{\partial \kappa_{\theta}}\mathcal{N}_{W_{\rm S}}}{(\mathcal{D})^2}.
\end{equation}

We first show the following Lemma:
\begin{lemma}
\label{lemmath2}
Consider $\mathcal{N}_{W_{\rm S}}$ and $\mathcal{D}$ as previously defined,
the following can be shown:
\begin{equation}
\partial\mathcal{N}_{W_{\rm S}}/\partial \kappa_{\theta}{>}\partial \mathcal{D}/\partial \kappa_{\theta},
\end{equation}
with
$0{\leq}{\theta}{\leq}T$.

\begin{proof}
If ${\theta}{=}T$, we have
\begin{equation}
\partial\mathcal{N}_{W_{\rm S}}/\partial \kappa_{T}{-}\partial \mathcal{D}/\partial \kappa_{T}{=}\alpha\prod_{i=1}^{T-1}(\rho{+}(1{-}\rho)\lambda\kappa_i)
\end{equation}
that is clearly positive.

Assume ${\theta}{<}T$. The expressions
\begin{equation}
\label{fpart}
1{+}(1{-}\rho)\lambda\kappa_{\theta+1}{+}\!\!\sum_{t{=}{\theta}+2}^{T}(1{-}\rho)\lambda\kappa_t\!\prod_{i{=}\theta+1}^{t-1}(\rho{+}(1{-}\rho)\lambda\kappa_i),
\end{equation}
and
\begin{equation}
\label{spart}
(1{-}\rho)\lambda\bigg(1{+}\sum_{t{=}{\theta}+1}^{T-1}\prod_{i={\theta}+1}^t(\rho{+}(1-\rho)\lambda\kappa_i)\bigg),
\end{equation}
are $\partial\mathcal{N}_{W_{\rm S}}/\partial \kappa_{\theta}$ and $\partial \mathcal{D}/\partial \kappa_{\theta}$ divided by 
$\alpha\prod_{i=1}^{{\theta}-1}(\rho{+}(1{-}\rho)\lambda\kappa_i)$, respectively.

Eqs.~(\ref{fpart}) and~(\ref{spart}) can be reorganized as 
\begin{equation}
\label{fpart2}
1{+}(1{-}\rho)\lambda\sum_{t{=}{\theta}+1}^{T}\rho^{t-{\theta}-1}\kappa_t{+}{C_1}.
\end{equation}
and
\begin{equation}
\label{spart2}
(1{-}\rho^{t-{\theta}-1})\lambda{+}(1{-}\rho)\lambda^2\sum_{t{=}{\theta}+1}^{T-1}(\rho^{t-{\theta}-1}{-}\rho^{t-{\theta}})\kappa_t{+}{C_2}.
\end{equation}
respectively, where the constants ${C_1}$ and ${C_2}$ account for all the cross-terms involving the multiplications of two or more
variables $\kappa_i$. We do not provide here the expressions for ${C_1}$ and ${C_2}$, as they
are conceptually simple, but tedious. However, it is possible to show that ${C_1}{>}{C_2}$.

Since $1{>}(1{-}\rho^{t-{\theta}-1})\lambda$ and
\begin{equation} 
(1{-}\rho)\lambda\!\!\!\sum_{t{=}{\theta}+1}^{T}\rho^{t-{\theta}-1}\kappa_t{>}(1{-}\rho)\lambda^2\!\!\!\sum_{t{=}{\theta}+1}^{T-1}(\rho^{t-{\theta}-1}{-}\rho^{t-{\theta}})\kappa_t,
\end{equation}
then $\partial\mathcal{N}_{W_{\rm S}}/\partial \kappa_{\theta}{>}\partial \mathcal{D}/\partial \kappa_{\theta}$.
\end{proof}
\end{lemma}

As for the proof of Theorem~\ref{th1}, since $(\mathcal{D})^2{>}0$, we focus on $\frac{\partial \mathcal{N}_{W_{\rm S}} }{\partial \kappa_{\theta}}\mathcal{D}{-}\frac{\partial \mathcal{D}}{\partial \kappa_{\theta}}\mathcal{N}_{W_{\rm S}}$. Due to Lemma~\ref{lemmath2}, 
$\frac{\partial \mathcal{N}_{W_{\rm S}} }{\partial \kappa_{\theta}}{>}  \frac{\partial \mathcal{D}}{\partial \kappa_{\theta}}$. Moreover,
$\mathcal{D}{\geq}\mathcal{N}_{W_{\rm S}}$. In fact,
\begin{equation}
\mathcal{W}_{\rm S}(\underline{\kappa})=\frac{\mathcal{N}_{W_{\rm S}}}{\mathcal{D}}\leq 1.
\end{equation}
Therefore,
\begin{equation}
\frac{\partial \mathcal{N}_{W_{\rm S}} }{\partial \kappa_{\theta}}\mathcal{D}>\frac{\partial \mathcal{D} }{\partial \kappa_{\theta}}\mathcal{D}\geq
\frac{\partial \mathcal{D} }{\partial \kappa_{\theta}}\mathcal{N}_{W_{\rm S}},
\end{equation}
and $\frac{\partial \mathcal{N}_{W_{\rm S}} }{\partial \kappa_{\theta}}\mathcal{D}{-}\frac{\partial \mathcal{D}}{\partial \kappa_{\theta}}\mathcal{N}_{W_{\rm S}}{>}0$.
\end{proof}
\begin{figure*}[!t]
\normalsize
\setcounter{mytempeqncnt}{\value{equation}}
\setcounter{equation}{83}
\begin{equation}
\label{Jj}
\mathcal{J}(\underline{\kappa}{+}\underline{u}_j\delta_j^{\prime})=
\frac{\mathcal{N}_J(\underline{\kappa})+\Delta\mathcal{N}_{J}(j,\delta^{\prime}_j,\underline{\kappa})+\Delta\mathcal{N}_{J}(r,\delta^{\prime}_j,\underline{\kappa})-\Delta\mathcal{N}_{J}(r,\delta^{\prime}_j,\underline{\kappa})  }{\mathcal{D}(\underline{\kappa})+\Delta\mathcal{D}(j,\delta^{\prime}_j,\underline{\kappa})+\Delta\mathcal{D}(r,\delta^{\prime}_j,\underline{\kappa})-\Delta\mathcal{D}(r,\delta^{\prime}_j,\underline{\kappa})}=\frac{\mathcal{N}_J+ (B+C)~ \delta_j^{\prime}  }{\mathcal{D}+(A+C)~\delta_j^{\prime}}
\end{equation}
\begin{equation}
\label{Jr}
\mathcal{J}(\underline{\kappa}{+}\underline{u}_r\delta_r^{\prime\prime})=
\frac{\mathcal{N}_J(\underline{\kappa})+ \Delta\mathcal{N}_{J}(r,\delta^{\prime\prime}_r,\underline{\kappa}) }{\mathcal{D}(\underline{\kappa})+\Delta\mathcal{D}(r,\delta^{\prime\prime}_r,\underline{\kappa})}
=\frac{\mathcal{N}_J +B~ \delta_r^{\prime\prime}  }{\mathcal{D}+A~\delta_r^{\prime\prime}}
\end{equation}
\begin{equation}
\label{Wj}
\mathcal{W}(\underline{\kappa}{+}\underline{u}_j\delta_j^{\prime})=\frac{\mathcal{N}_W(\underline{\kappa})+\Delta\mathcal{N}_{W}(j,\delta^{\prime}_j,\underline{\kappa})+\Delta\mathcal{N}_{W}(r,\delta^{\prime}_j,\underline{\kappa})-\Delta\mathcal{N}_{W}(r,\delta^{\prime}_j,\underline{\kappa}) }{\mathcal{D}(\underline{\kappa})+\Delta\mathcal{D}(j,\delta^{\prime}_j,\underline{\kappa})+\Delta\mathcal{D}(r,\delta^{\prime}_j,\underline{\kappa})-\Delta\mathcal{D}(r,\delta^{\prime}_j,\underline{\kappa})}=\frac{\mathcal{N}_W(\underline{\kappa})+ (G+F)~ \delta_j^{\prime}  }{\mathcal{D}+(A+C)~\delta_j^{\prime}}
\end{equation}
\begin{equation}
\label{Wr}
\mathcal{W}(\underline{\kappa}{+}\underline{u}_r\delta^{\prime\prime}_r)=\frac{\mathcal{N}_W(\underline{\kappa})+ \Delta\mathcal{N}_{W}(r,\delta^{\prime\prime}_r,\underline{\kappa}) }{\mathcal{D}(\underline{\kappa})+\Delta\mathcal{D}(r,\delta^{\prime\prime}_r,\underline{\kappa})}=\frac{\mathcal{N}_W+G~ \delta_r^{\prime\prime}  }{\mathcal{D}+A~\delta_r^{\prime\prime}}.
\end{equation}
\hrulefill
\vspace*{1pt}
\setcounter{equation}{\value{mytempeqncnt}}
\vup\vup\vup
\end{figure*}

\section{Proof of Theorem~\ref{mainth}}
\label{appmth}
\begin{proof}
Theorem~\ref{mainth} states that starting from a policy $\underline{\kappa}$, for any pair of indices $j$ and $r$ with $0{<}j{<}r$
such that $\kappa_j{=}\kappa_r$ and $\kappa_t{=}0$, $r{<}t{\leq}T$, if 
\begin{equation}
\mathcal{J}_{\rm P}(\underline{\kappa}^{\prime}){=}\mathcal{J}_{\rm P}(\underline{\kappa}^{\prime\prime}),
\end{equation}
then
\begin{equation}
\mathcal{W}_{\rm S}(\underline{\kappa}^{\prime}){>}\mathcal{W}_{\rm S}(\underline{\kappa}^{\prime\prime}),
\end{equation}
with $\underline{\kappa}^{\prime}{=}\underline{\kappa}{+}\underline{u}_j\delta_j^{\prime}$ and $\underline{\kappa}^{\prime\prime}{=}\underline{\kappa}{+}\underline{u}_r\delta_r^{\prime\prime}$, $0{<}\delta_j^{\prime}{\leq} 1{-}\kappa_j$ and $0{<}\delta_r^{\prime\prime}{\leq}1{-}\kappa_r$.

In words, if the policy obtained by increasing the $j$--th element of $\underline{\kappa}$ by $\delta^{\prime}_j$
and the policy obtained by increasing the  $r$--th element of $\underline{\kappa}$ by $\delta_r^{\prime\prime}$ incur the same primary source's average cost, then the former policy achieves a larger secondary source's average reward.

We briefly recall the notation introduced in the previous proofs. The average primary source's cost and secondary source's reward can be written
respectively as 
\begin{eqnarray}
\mathcal{J}_{\rm P}(\underline{\kappa})\!&=&\!\frac{\mathcal{N}_{J_{\rm P}}(\underline{\kappa})}{\mathcal{D}(\underline{\kappa})},\\
\mathcal{W}_{\rm S}(\underline{\kappa})\!&=&\!\frac{\mathcal{N}_{W_{\rm S}}(\underline{\kappa})}{\mathcal{D}(\underline{\kappa})}.
\end{eqnarray}
where
\begin{eqnarray}
\mathcal{N}_{J_{\rm P}}(\underline{\kappa})\!&\!\!\!{=}\!\!\!&\!(1{-}\alpha){+}\alpha\sum_{t=1}^{T}\prod_{i=1}^{t}(\rho{+} (1 {-} \rho)\lambda \kappa_i){>}0\\
\mathcal{D}(\underline{\kappa})\!&\!\!\!{=}\!\!\!&\!1{+}\alpha\sum_{t=1}^{T-1}\prod_{i=1}^{t}(\rho{+} (1 {-} \rho)\lambda \kappa_i){>}0,\\
\mathcal{N}_{W_{\rm S}}(\underline{\kappa})\!&\!\!\!{=}\!\!\!&\!(1{-}\alpha){+}\alpha\sum_{t=1}^{T}\kappa_{t}\prod_{i=1}^{t-1}(\rho{+} (1 {-} \rho)\lambda \kappa_i){>}0
\end{eqnarray}

In the following, in order to simplify the notation, we drop the subscripts ${\rm P}$ and ${\rm S}$ and we refer to the primary source's
cost and secondary source's reward when talking of cost and reward, respectively.

If the $q$--th element of $\underline{\kappa}$, with $q{>}0$, is increased by $\delta$, with $\delta{\leq}1{-}\kappa_q$, the average cost and reward can be written as
\begin{eqnarray}
\mathcal{J}(\underline{\kappa}{+}\underline{u}_q \delta)\!&=&\!\frac{\mathcal{N}_{J}(\underline{\kappa}){+}\Delta\mathcal{N}_{J}(q,\delta,\underline{\kappa})}{\mathcal{D}(\underline{\kappa}){+}\Delta\mathcal{D}(q,\delta,\underline{\kappa})},\\
\mathcal{W}(\underline{\kappa}{+}\underline{u}_q \delta)\!&=&\!\frac{\mathcal{N}_{W}(\underline{\kappa}){+}\Delta\mathcal{N}_{W}(q,\delta,\underline{\kappa})}{\mathcal{D}(\underline{\kappa}){+}\Delta\mathcal{D}(q,\delta,\underline{\kappa})}.
\end{eqnarray}
where
\begin{eqnarray}
\!\!\!\Delta\mathcal{N}_{J}(q,\delta,\underline{\kappa})\!\!&\!\!\!{=}\!\!\!&\!\! \delta \bigg[\alpha(1{-}\rho)\lambda \sum_{t=q}^{T}\prod_{i=1,i\neq q}^{t}\!\!\!(\rho{+}(1{-}\rho)\lambda\kappa_i)\bigg]\!,\\
\!\!\!\Delta\mathcal{D}(q,\delta,\underline{\kappa})\!\!&\!\!\!{=}\!\!\!&\!\! \delta\bigg[ \alpha(1{-}\rho)\lambda \sum_{t=q}^{T-1}\prod_{i=1,i\neq q}^{t}\!\!\!(\rho{+}(1{-}\rho)\lambda\kappa_i)\bigg]\!,\\
\!\!\!\Delta\mathcal{N}_{W}(q,\delta,\underline{\kappa})\!\!&\!\!\!{=}\!\!\!&\!\!\delta\bigg[\alpha\prod_{i=1}^{q-1}(\rho{+}(1{-}\rho)\lambda\kappa_i){+}\nonumber\\
{+}\!\!&\!\!\!{\alpha}\!\!\!&\!\!(1{-}\rho)\lambda \!\! \sum_{t=q+1}^{T}\!\!\kappa_t\!\!\prod_{i=1,i\neq q}^{t-1}\!\!\!(\rho{+}(1{-}\rho)\lambda\kappa_i)\bigg]
\end{eqnarray}
\begin{figure*}[!t]
\normalsize
\setcounter{mytempeqncnt}{\value{equation}}
\setcounter{equation}{91}
\begin{equation}
\label{equopt2}
\frac{(\mathcal{D} Z - \mathcal{N}_J)(\mathcal{D} (B F {-} C G) {+} (C G{-}A F) \mathcal{N}_J {+} (A {-} B) C \mathcal{N}_W)}{(B \mathcal{D} - A \mathcal{N}_J) ((B + C) \mathcal{D} - (A + C) \mathcal{N}_J)}>0.
\end{equation}
\setcounter{equation}{\value{mytempeqncnt}}
\vup\vup\vup
\hrulefill
\vspace*{1pt}
\end{figure*}

Thus, $\Delta\mathcal{N}_{J}(q,\delta,\underline{\kappa})$, $\Delta\mathcal{N}_{W}(q,\delta,\underline{\kappa})$ and $\Delta\mathcal{D}(q,\delta,\underline{\kappa})$ are linear functions of $\delta$,
and represent the increment of the numerator of the cost and reward, and of their denominator, corresponding to an increase $\delta$ of $\kappa_q$. Note
that if $\delta$ is strictly positive and $q{>}0$,\footnote{Together with the assumptions $\lambda,\alpha,1{-}\rho{>}0$.} then $\Delta\mathcal{N}_{J}(q,\delta,\underline{\kappa})$, $\Delta\mathcal{D}(q,\delta,\underline{\kappa})$ and $\Delta\mathcal{N}_{W}(q,\delta,\underline{\kappa})$
are strictly positive.

According to the hypothesis of the theorem $\mathcal{J}(\underline{\kappa}{+}\underline{u}_j\delta^{\prime}_j){=}\mathcal{J}(\underline{\kappa}{+}\underline{u}_r\delta^{\prime\prime}_r)$. This equality can be rewritten as 
\begin{equation}
\label{equality1}
\frac{\mathcal{N}_J(\underline{\kappa}){+}\Delta\mathcal{N}_{J}(j,\delta_j^{\prime},\underline{\kappa})}{\mathcal{D}(\underline{\kappa}){+}\Delta\mathcal{D}(j,\delta^{\prime}_j,\underline{\kappa})}=\frac{\mathcal{N}_J(\underline{\kappa}){+}\Delta\mathcal{N}_{J}(r,\delta_r^{\prime\prime},\underline{\kappa})}{\mathcal{D}(\underline{\kappa}){+}\Delta\mathcal{D}(r,\delta_r^{\prime\prime},\underline{\kappa})}.
\end{equation}

The increases of the numerators and denominators can be rewritten as 
\begin{eqnarray}
\label{lin1}
\Delta\mathcal{D}(r,\delta,\underline{\kappa})&=&\delta A,\\
\label{lin2}
\Delta\mathcal{N}_J(r,\delta,\underline{\kappa})&=&\delta B,\\
\label{lin3}
\Delta\mathcal{N}_W(r,\delta,\underline{\kappa})&{=}&\delta G.
\end{eqnarray}
Note that,
since $\kappa_j{=}\kappa_r$, the difference between
the increase of the denominator when $\kappa_j$ or $\kappa_r$ are increased by $\delta$ is a constant $C$ equal to
\begin{eqnarray}
\label{lin4}
\Delta\mathcal{D}(j,\delta,\underline{\kappa})\!\!&\!\!\!{-}\!\!\!&\!\!\Delta\mathcal{D}(r,\delta,\underline{\kappa})=\nonumber\\
\!\!&\!\!\!{=}\!\!\!&\!\!\delta\alpha(1{-}\rho)\lambda\sum_{t{=}j}^{r-1}\prod_{i{=}1,i{\neq}j}^t\!\!(\rho{+}(1{-}\rho)\lambda\kappa_i)\nonumber\\
\!\!&\!\!\!{=}\!\!\!&\!\!\delta C>0.
\end{eqnarray}

Analogously, the difference between the numerators of the cost and reward increases are
\begin{eqnarray}
\label{lin5}
\Delta\mathcal{N}_{J}(j,\delta,\underline{\kappa})-\Delta\mathcal{N}_{J}(r,\delta,\underline{\kappa})\!\!&\!\!\!{=}\!\!\!&\!\!\Delta\mathcal{D}(j,\delta,\underline{\kappa}){-}\Delta\mathcal{D}(r,\delta,\underline{\kappa})\nonumber\\
\!\!&\!\!\!{=}\!\!\!&\!\!\delta C,
\end{eqnarray}
and
\begin{eqnarray}
\label{lin6}
\!\!&\!\!\!{\Delta}\!\!\!&\!\!\mathcal{N}_{W}(j,\delta,\underline{\kappa})-\Delta\mathcal{N}_{W}(r,\delta,\underline{\kappa}){=}\nonumber\\
\!\!&\!\!\!{=}\!\!\!&\!\!\delta\alpha\bigg[
\prod_{i=1}^{j-1}(\rho{+}(1{-}\rho)\lambda\kappa_i)-\prod_{i=1}^{r-1}(\rho+(1{-}\rho)\lambda\kappa_i)\bigg]{+}\nonumber\\
\!\!&\!\!\!{+}\!\!\!&\!\!\delta\alpha(1{-}\rho)\lambda\sum_{t=j+1}^r\kappa_t\prod_{i=1,i\neq j}^{t-1}(\rho{+}(1{-}\rho)\lambda\kappa_i)\nonumber\\
\!\!&\!\!\!{=}\!\!\!&\!\! \delta F>0,
\end{eqnarray}
respectively.

According to Eqs.~(\ref{lin1})-(\ref{lin6}), and omitting the dependency of the quantities on $\underline{\kappa}$, we rewrite
$\mathcal{J}(\underline{\kappa}{+}\underline{u}_j\delta_j^{\prime})$, $\mathcal{J}(\underline{\kappa}{+}\underline{u}_r\delta_r^{\prime\prime})$,
$\mathcal{W}(\underline{\kappa}{+}\underline{u}_j\delta_j^{\prime})$ and $\mathcal{W}(\underline{\kappa}{+}\underline{u}_r\delta_r^{\prime\prime})$
as shown in Eqs.~(\ref{Jj}),~(\ref{Jr}),~(\ref{Wj}) and~(\ref{Wr}), respectively.\addtocounter{equation}{4}

Note that
\begin{equation}
\mathcal{J}(\underline{\kappa}{+}\underline{u}_j\delta){=}\frac{\mathcal{N}_J{+} (B{+}C)~ \delta  }{\mathcal{D}{+}(A{+}C)~\delta}{>}
\frac{\mathcal{N}_J{+} B~ \delta  }{\mathcal{D}{+}A~\delta}{=}\mathcal{J}(\underline{\kappa}{+}\underline{u}_r\delta)
\end{equation}
for any $\delta$, with $0{<}\delta{\leq}\min(1{-}\kappa_j,1{-}\kappa_r)$.

Choose $\delta^{\prime\prime}_r$, with $0{<}\delta^{\prime\prime}_r{\leq}1{-}\kappa_r$, and
denote the cost of policy $\underline{\kappa}{+}\underline{u}_r\delta^{\prime\prime}_{r}$ with 
$Z{=}\mathcal{J}(\underline{\kappa}{+}\underline{u}_r\delta^{\prime\prime}_{r})$. Observe that,
due to the monotonicity of the cost function, then $\mathcal{J}(\underline{\kappa}){<}Z{\leq}1$.

Since the cost function is continuous with respect to any element of the policy vector 
and $\kappa_j{=}\kappa_r$ by assumption, then there always exists $\delta_j^{\prime}$
such that $\mathcal{J}(\underline{\kappa}{+}\underline{u}_j\delta^{\prime}_{j}){=}\mathcal{J}(\underline{\kappa}{+}\underline{u}_r\delta^{\prime\prime}_{r}){=}Z$, with $0{<}\delta^{\prime}_j{<}\delta^{\prime\prime}_r{\leq}1{-}\kappa_r{=}1{-}\kappa_j$.

The values for $\delta^{\prime}_j$ and $\delta^{\prime\prime}_r$ can be readily found to be
\begin{eqnarray}
\label{deltaval1}
\delta_j^{\prime}\!\!&\!\!\!{=}\!\!\!&\!\!  \frac{\mathcal{D} ~Z-\mathcal{N}_J}{B+C-(A+C) ~Z} \\
\label{deltaval2}
\delta_r^{\prime\prime}\!\!&\!\!\!{=}\!\!\!&\!\! \frac{\mathcal{D} ~Z-\mathcal{N}_J}{B-A ~Z}.
\end{eqnarray}

In order to complete the proof, the following inequality needs to be proved:
\begin{eqnarray}
\label{equopt}
\!\!&\!\!\mathcal{W}\!\!&\!\!(\underline{\kappa}{+}\underline{u}_j\delta_j^{\prime})-\mathcal{W}(\underline{\kappa}{+}\underline{u}_r\delta_r^{\prime\prime}){=}\nonumber\\
\!\!&\!\!{=}\!\!&\!\!\frac{\mathcal{N}_W{+} (G{+}F)~ \delta_j^{\prime}  }{\mathcal{D}{+}(A{+}C)~\delta_j^{\prime}}{-}\frac{\mathcal{N}_W{+}G~ \delta_r^{\prime\prime}  }{\mathcal{D}{+}A~\delta_r^{\prime\prime}}>0
\end{eqnarray}

By substituting Eq.~(\ref{deltaval1}) and Eq.~(\ref{deltaval2}) in Eq.~(\ref{equopt}), we obtain Eq.~(\ref{equopt2}).\addtocounter{equation}{1} Note that 
\begin{equation}
\frac{\mathcal{D} Z - \mathcal{N}_J}{(B \mathcal{D} - A \mathcal{N}_J) ((B + C) \mathcal{D} - (A + C) \mathcal{N}_J)}{>}0.
\end{equation}

In fact, recalling that $\kappa_j{=}\kappa_r$ by hypothesis, we have 
\begin{eqnarray}
B-A\!\!&\!\!\!{=}\!\!\!&\!\!\alpha(1{-}\rho)\lambda\bigg(\prod_{i=1,i\neq r}^{T}(\rho{+}(1-\rho)\lambda)\kappa_i\bigg)\nonumber\\
\!\!&\!\!\!{=}\!\!\!&\!\!\alpha(1{-}\rho)\lambda\bigg(\prod_{i=1,i\neq j}^{T}(\rho{+}(1-\rho)\lambda)\kappa_i\bigg)>0
\end{eqnarray}
and since $\mathcal{D}\geq\mathcal{N}_J$,\footnote{Because $\frac{\mathcal{N}_J}{\mathcal{D}}{=}\mathcal{J}(\underline{\kappa}){\leq} 1$.}
then
\begin{equation}
\label{eqfuffa}
(B \mathcal{D} - A \mathcal{N}_J) ((B + C) \mathcal{D} - (A + C) \mathcal{N}_J){>}0.
\end{equation}
Moreover, due to Theorem~\ref{th1}, we have for $\delta^{\prime}_j{>}0$
\begin{equation}
Z{=}\mathcal{J}(\underline{\kappa}{+}\underline{u}_j\delta_j^{\prime}){>}\mathcal{J}(\underline{\kappa}){=}\frac{\mathcal{N}_J }{\mathcal{D}}.
\end{equation}
Therefore,
\begin{equation}
\mathcal{D} Z - \mathcal{N}_J >0.
\end{equation}
\begin{figure*}[!t]
\normalsize
\setcounter{mytempeqncnt}{\value{equation}}
\setcounter{equation}{102}
\begin{eqnarray}
(\mathcal{D}{-}\mathcal{N}_J)F~B {+} (B-A) \mathcal{N}_J F \!\!&\!\!\!{=}\!\!\!&\!\!\alpha F ~B - X~ F~ B + \frac{(1-\rho)\lambda}{\rho+(1-\rho)\lambda\kappa_j} \mathcal{N}_J\nonumber\\\!\!&\!\!\!{=}\!\!\!&\!\! \alpha~ F ~B + (1{-}\alpha) \frac{(1-\rho)\lambda}{\rho+(1-\rho)\lambda\kappa_j} X ~F - X~F \alpha (1-\rho)\lambda ~X~F \prod_{i=1,i\neq j}^T(\rho+(1-\rho)\lambda\kappa_i)\nonumber\\
\label{termsempl1}\!\!&\!\!\!{}\!\!\!&\!\!+ X ~ F \alpha (1-\rho)\lambda \sum_{t=1}^{J-1}\prod_{i=1,i\neq j}^t (\rho+(1-\rho)\lambda \kappa_i) + X~F~C.
\end{eqnarray}
\setcounter{equation}{104}
\begin{eqnarray}
{-}(\mathcal{D}{-}\mathcal{N}_J)G~C {-} (B-A) \mathcal{N}_W C \!\!&\!\!\!{=}\!\!\!&\!\! -\alpha~ C~G {+}X~C ~G -  \frac{(1-\rho)\lambda}{\rho+(1-\rho)\lambda\kappa_j} X ~C ~\mathcal{N}_W\nonumber\\ 
\!\!&\!\!\!{=}\!\!\!&\!\! - \alpha~C~G - (1{-}\alpha) \frac{(1-\rho)\lambda}{\rho+(1-\rho)\lambda\kappa_j} X ~C{+}X~C~\alpha\prod_{i=1}^{r-1}(\rho{+}(1-\rho)\lambda\kappa_i)+\nonumber\\
\!\!&\!\!\!{}\!\!\!&\!\! -X~C~\alpha(1-\rho)\lambda\sum_{t=1}^r\kappa_t\prod_{i=1,i\neq j}^{t-1}(\rho{+}(1-\rho)\lambda\kappa_i)\nonumber\\
\!\!&\!\!\!{=}\!\!\!&\!\! -\alpha~C~G{-} \frac{(1-\rho)\lambda}{\rho+(1-\rho)\lambda\kappa_j} X ~ (1{-}\alpha)~C + X ~C \alpha(1-\rho)\lambda\sum_{t=1}^j\kappa_t
\prod_{i=1,i\neq j}^{t-1}(\rho+(1-\rho)\lambda k_i)+\nonumber\\
\label{termsempl2}
\!\!&\!\!\!{}\!\!\!&\!\!-X~C~F + X~C~\alpha\prod_{i=1}^{j-1}(\rho+(1-\rho)\lambda\kappa_i).
\end{eqnarray}
\begin{eqnarray}
\mathcal{D} (B F \!\!&\!\!\!{-}\!\!\!&\!\! C G) {+} (C G{-}A F) \mathcal{N}_J {+} (A {-} B) C \mathcal{N}_W = \alpha~ F \bigg( B{-}X ~\alpha(1-\rho)\lambda\prod_{i=1,i\neq j}^T(\rho{+}(1-\rho)\lambda\kappa_i)  ~\bigg)+\nonumber\\
\!\!&\!\!\!{}\!\!\!&\!\!(1-\alpha)\frac{(1-\rho)\lambda}{\rho+(1-\rho)\lambda\kappa_j}X~(F-C)
+ X~F~\alpha(1-\rho)\lambda \sum_{t=1}^{j-1}\prod_{i=1,i\neq j}^t(\rho+(1-\rho)\lambda\kappa_i) +\nonumber\\\label{tutto}\!\!&\!\!\!{}\!\!\!&\!\!
 X~C~\alpha\prod_{i=1}^{j-1}(\rho+(1-\rho)\lambda\kappa_i)+
X~C~\alpha(1-\rho)\lambda\sum_{t=1}^j\kappa_t\prod_{i=1,i\neq j}^{t-1}(\rho+(1-\rho)\lambda \kappa_i)
\end{eqnarray}
\setcounter{equation}{\value{mytempeqncnt}}
\hrulefill
\vspace*{1pt}
\vup
\end{figure*}

The Theorem is then proved if the following inequality holds: 
\begin{eqnarray}
\mathcal{D} (B F \!\!&\!\!\!{-}\!\!\!&\!\! C G) {+} (C G{-}A F) \mathcal{N}_J {+} (A {-} B) C \mathcal{N}_W){=}\nonumber\\
\label{esprl1}
(\mathcal{D}{-}\mathcal{N}_J\!\!&\!\!\!)(\!\!\!&\!\!F~B {-} C~G){+}(B{-}A)(\mathcal{N}_J ~F{-}\mathcal{N}_W ~C){>}0.
\end{eqnarray}

Define
\begin{equation}
X=\alpha\prod_{i=1}^T (\rho+(1-\rho)\lambda \kappa_i)>0.
\end{equation}
Then,
\begin{equation}
\mathcal{D}{-}\mathcal{N}_J{=}\alpha\bigg(1-\prod_{i=1}^{T}(\rho{+}(1{-}\rho)\lambda\kappa_i)\bigg)=\alpha{-}X>0,
\end{equation}
and 
\begin{equation}
B{-}A=\frac{(1-\rho)\lambda}{\rho+(1-\rho)\lambda\kappa_j}X>0.
\end{equation}

In the second term of Eq.~(\ref{esprl1}), $\mathcal{N}_J$ can be split into two terms
\begin{eqnarray}
\mathcal{N}_J \!\!&\!\!\!=\!\!\!&\!\! (1{-}\alpha){+}\alpha\sum_{t=1}^{r-1}\prod_{i=1}^t (\rho+(1-\rho)\lambda\kappa_i)+\nonumber\\
\!\!&\!\!\!+\!\!\!&\!\!\alpha\sum_{t=r}^{T-1}\prod_{i=1}^t (\rho+(1-\rho)\lambda\kappa_i),
\end{eqnarray}
so that the second summation corresponds to the summation in $B$.\footnote{In $B$, the summation has the additional term corresponding to $t{=}T$ and 
the products do not have the term $i{=}j$.} $(\mathcal{D}{-}\mathcal{N}_J)F~B {+} (B-A) \mathcal{N}_J F$ 
in Eq.~(\ref{esprl1}) can be simplified as shown in Eq.~(\ref{termsempl1}).\addtocounter{equation}{1}

Analogously, $\mathcal{N}_W$ can be rewritten as
\begin{eqnarray}
\mathcal{N}_W \!\!&\!\!\!=\!\!\!&\!\! (1{-}\alpha){+}\alpha\sum_{t=1}^{r}\kappa_t \prod_{i=1}^{t-1}(\rho+(1-\rho)\lambda\kappa_i)+\nonumber\\
\!\!&\!\!\!+\!\!\!&\!\!\alpha\sum_{t=r+1}^{T}\kappa_t\prod_{i=1}^{t-1} (\rho+(1-\rho)\lambda\kappa_i),
\end{eqnarray}
so that one summation corresponds to the summation in $G$. The term  $-(\mathcal{D}{-}\mathcal{N}_J)G~C {-} (B-A) \mathcal{N}_W C$ 
in Eq.~(\ref{esprl1}) can be rewritten as reported in Eq.~(\ref{termsempl2}).\addtocounter{equation}{2}

Eq.~(\ref{esprl1}) is the sum of Eq.~(\ref{termsempl1}) and Eq.~(\ref{termsempl2}).
By hypothesis $\kappa_t{=}0$, $\forall t{>}r$, and thus $G{=}0$. The term $-\alpha~G~C$
in Eq.~(\ref{termsempl2}) is then equal to zero. Eq.~(\ref{tutto}) reorganizes the sum of  
Eqs.~(\ref{termsempl1}) and (\ref{termsempl2}).

The first term of Eq.~(\ref{tutto})
\begin{equation}
\alpha~ F \bigg( B{-}X ~\alpha(1-\rho)\lambda\prod_{i=1,i\neq j}^T(\rho{+}(1-\rho)\lambda\kappa_i)  ~\bigg)
\end{equation}
is positive. In fact, $X{\leq}1$ and 
\begin{eqnarray}
B\!\!&\!\!\!-\!\!\!&\!\! \alpha(1-\rho)\lambda\prod_{i=1,i\neq j}^T(\rho{+}(1-\rho)\lambda\kappa_i)  ~\bigg){=}\nonumber\\
\!\!&\!\!\!=\!\!\!&\!\! \alpha(1-\rho)\lambda\sum_{t=r}^{T-1} \prod_{i=1,i\neq j}^t(\rho{+}(1-\rho)\lambda\kappa_i)>0
\end{eqnarray}

Moreover, it can be shown that $F{>}C$. The proof is analogous to that of Lemma~\ref{lemmath2},
and is not reported herein. As a consequence, the second term of Eq.~(\ref{tutto}) is positive.
All the other terms are trivially positive.

The inequality is then proved, as well as the Theorem.
\end{proof}

\section{Proof of Theorem~\ref{mainth2}}
\label{appmth2}
\begin{proof}
Theorem~\ref{mainth2} states that starting from a policy $\underline{\kappa}$, for any pair of indices $j$ and $r$, with $0{<}j{<}r$,
such that $\kappa_j{=}\kappa_r$ and $\kappa_t{=}0$, $r{<}t{\leq}T$, if 
\begin{equation}
\mathcal{J}_{\rm P}(\underline{\kappa}^{\prime}){=}\mathcal{J}_{\rm P}(\underline{\kappa}^{\prime\prime}),
\end{equation}
then
\begin{equation}
\mathcal{W}_{\rm S}(\underline{\kappa}^{\prime}){<}\mathcal{W}_{\rm S}(\underline{\kappa}^{\prime\prime}),
\end{equation}
with $\underline{\kappa}^{\prime}{=}\underline{\kappa}{-}\underline{u}_j\delta_j^{\prime}$ and $\underline{\kappa}^{\prime\prime}{=}\underline{\kappa}{-}\underline{u}_r\delta_r^{\prime\prime}$, $0{<}\delta_j^{\prime}{\leq} \kappa_j$ and $0{<}\delta_r^{\prime\prime}{\leq}\kappa_r$.
The proof is similar to that provided in the previous Appendix. 

As done in the previous proof, we fix $\delta_r^{\prime\prime}$, with $0{<}\delta_r^{\prime\prime}{\leq}1{-}\kappa_r$,
and we denote the cost associated with the policy obtained by decreasing $\kappa_r$ by $\delta^{\prime\prime}_r$ with $Z{=}\mathcal{J}(\underline{\kappa}{-}\underline{u}_r\delta^{\prime\prime}_r)$. Note that $0{\leq}Z{<}\mathcal{J}_{\rm P}(\underline{\kappa})$. Through considerations entirely analogous
to those provided in Appendix~\ref{appmth}, it can be shown that there exists $\delta^{\prime}_j$ such that $\mathcal{J}_{\rm P}(\underline{\kappa}{-}\underline{u}_j\delta_j^{\prime}){=} \mathcal{J}_{\rm P}(\underline{\kappa}{-}\underline{u}_r\delta_r^{\prime\prime}){=}Z$, with $0{<}\delta^{\prime}_j{<}\delta^{\prime\prime}_r{\leq}\kappa_r{=}\kappa_j$.
The corresponding values of $\delta^{\prime}_j$ and $\delta^{\prime\prime}_r$ can be readily found to be
\begin{eqnarray}
\label{deltaval1th2}
\delta_j^{\prime}\!\!&\!\!{=}\!\!&\!\!  \frac{\mathcal{N}_J - \mathcal{D} ~Z}{B+C-(A+C) ~Z} \\
\label{deltaval2th2}
\delta_r^{\prime\prime}\!\!&\!\!{=}\!\!&\!\! \frac{\mathcal{N}_J-\mathcal{D} ~Z}{B-A ~Z}.
\end{eqnarray}
Observe that the above values of $\delta^{\prime}_j$ and $\delta_r^{\prime\prime}$ are the opposites of those in Eqs.~(\ref{deltaval1}) and~(\ref{deltaval2}).

By substituting the above equations in
\begin{equation}
\mathcal{W}_{\rm S}(\underline{\kappa}^{\prime}){-}\mathcal{W}_{\rm S}(\underline{\kappa}^{\prime\prime}){=}\frac{\mathcal{N}_W{-} (G{+}F)~ \delta_j^{\prime}  }{\mathcal{D}{-}(A{+}C)~\delta_j^{\prime}}{-}\frac{\mathcal{N}_W{-} G~ \delta_r^{\prime\prime}  }{\mathcal{D}{-}A~\delta_r^{\prime\prime}},
\end{equation}
the same fraction as in Eq.~(\ref{equopt2}) is obtained, and
according to Eqs.~(\ref{eqfuffa}) and~(\ref{esprl1}),
\begin{equation}
\frac{(\mathcal{D} (B F {-} C G) {+} (C G{-}A F) \mathcal{N}_J {+} (A {-} B) C \mathcal{N}_W)}{(B \mathcal{D} - A \mathcal{N}_J) ((B + C) \mathcal{D} - (A + C) \mathcal{N}_J)}>0.
\end{equation}

Differently from the proof of Theorem~\ref{mainth}, since $Z{<}\mathcal{J}_{\rm P}(\underline{\kappa}){=}\mathcal{N}_J/\mathcal{D}$,
then
\begin{equation}
\mathcal{D} Z - \mathcal{N}_J<0.
\end{equation}
Therefore,
\begin{equation}
\mathcal{W}_{\rm S}(\underline{\kappa}^{\prime})-\mathcal{W}_{\rm S}(\underline{\kappa}^{\prime\prime})<0,
\end{equation}
and the theorem is proved.
\end{proof}

\section{Proof of Theorems~\ref{mainth} and~\ref{mainth2} for Constraint on the Failure Probability}
\label{appmf}
\begin{proof}
As discussed in Section~\ref{avfprob}, the cost increase/decrease induced by an increase/decrease of the transmission
probability in state $\theta{>}0$ does not depend on $\theta$. 

In this case, a transmission probability increased by $\delta$,\footnote{A decreased transmission probability corresponds to a negative $\delta$ in the following.} in state $j$ or $r$ results in the same increase of the average cost. More formally, 
fix $j$,$r$ and $\delta$, with $0{<}j{<}r{\leq}T$, $-\min(\kappa_j,\kappa_r){\leq}\delta{\leq}\min(1{-}\kappa_j,1{-}\kappa_r)$ and $\delta{\neq}0$, then
$\mathcal{J}^{\rm fp}_{\rm P}(\underline{\kappa}{+} \underline{u}_j\delta){=}\mathcal{J}^{\rm fp}_{\rm P}(\underline{\kappa}{+}\underline{u}_r\delta)$.

Therefore, starting from a policy $\underline{\kappa}$, and defining the policies $\underline{\kappa}^{\prime}{=}\underline{\kappa}{+} \underline{u}_j\delta_j^{\prime}$ and $\underline{\kappa}^{\prime\prime}{=}\underline{\kappa}{+} \underline{u}_r\delta_r^{\prime\prime}$, then
\begin{equation}
\mathcal{J}_{\rm P}^{\rm fp}(\underline{\kappa}^{\prime}){=}\mathcal{J}_{\rm P}^{\rm fp}(\underline{\kappa}^{\prime\prime}){=}Z,
\end{equation}
with $0{\leq}Z{\leq}1$, $Z{\neq}\mathcal{J}_{\rm P}^{\rm fp}(\underline{\kappa})$, only if $\delta^{\prime}_j{=}\delta^{\prime\prime}_r{=}\delta(Z)$, where $\delta(Z)$ is a function of $Z$.
Note that $\delta(Z){>}0$ if $Z{>}\mathcal{J}_{\rm P}^{\rm fp}(\underline{\kappa})$, and $\delta(Z){<}0$ otherwise.

According to the notation introduced in Appendix~\ref{appmth}, the difference between the rewards achieved with policies $\underline{\kappa}^{\prime}$ and $\underline{\kappa}^{\prime\prime}$ is
\begin{eqnarray}
\mathcal{W}_{\rm S}(\underline{\kappa}^{\prime})\!\!&\!\!\!-\!\!\!&\!\!\mathcal{W}_{\rm S}(\underline{\kappa}^{\prime\prime})=\nonumber\\
\!\!&\!\!\!{=}\!\!\!&\!\!\frac{\mathcal{N}_W{+} (G{+}F)~ \delta(Z)  }{\mathcal{D}{+}(A{+}C)~\delta(Z)}{-}\frac{\mathcal{N}_W{+} G~ \delta(Z)  }{\mathcal{D}{+}A~\delta(Z)},
\end{eqnarray}
which is larger than zero if the following holds:
\begin{equation}
\delta(Z)\bigg(F~(\mathcal{D}+A\delta(Z))-C( \mathcal{N}_W + G\delta(Z))\bigg)>0.
\end{equation}

Since $F{>}C$, as previously stated,
and
\begin{equation}
\mathcal{W}(\underline{\kappa}^{\prime\prime})=\mathcal{W}(\underline{\kappa}+\underline{u}_r \delta({Z})){=}\frac{ \mathcal{N}_W + G\delta(Z)}{\mathcal{D}+A\delta(Z)}\leq1
\end{equation}
by construction, then, 
\begin{equation}
F~(\mathcal{D}+A\delta(Z))-C( \mathcal{N}_W + G\delta(Z))>0.
\end{equation}

Finally, if $Z{>}\mathcal{J}_{\rm P}^{\rm fp}(\underline{\kappa})$, then $\delta(Z){>}0$
and
$\mathcal{W}(\underline{\kappa}^{\prime}){>}\mathcal{W}(\underline{\kappa}^{\prime\prime})$.
If $Z{<}\mathcal{J}_{\rm P}^{\rm fp}(\underline{\kappa})$, then $\delta(Z){<}0$ and
$\mathcal{W}(\underline{\kappa}^{\prime}){<}\mathcal{W}(\underline{\kappa}^{\prime\prime})$.
\end{proof}
\renewcommand{\baselinestretch}{0.9}
\bibliographystyle{IEEEtran}
\bibliography{IEEEabrv,cogbib_jitr}

\begin{biographynophoto}{Marco Levorato}(SÕ06, MÕ09) obtained both the 
BE (Electronics and Telecommunications Engineer- 
ing) and the ME (Telecommunications Engineering) 
summa cum laude from the University of Ferrara, 
Italy, in 2002 and 2005, respectively. In 2009, he 
received a Ph.D. in Information Engineering from 
the University of Padova. During 2008 he was on 
leave at the University of Southern California, Los 
Angeles, United States. In 2009 he was a post 
doctorate researcher at the University of Padova. 
Since January 2010, he is a post doctorate researcher 
at Stanford and the University of Southern California (USC). 
\end{biographynophoto}

\begin{biographynophoto}{Urbashi Mitra} (F'07)
Urbashi Mitra received the B.S. and the M.S. degrees from the University of California at Berkeley in 1987 and 1989 respectively, both in Electrical Engineering and Computer Science. From 1989 until 1990 she worked as a Member of Technical Staff at Bellcore in Red Bank, NJ.  In 1994, she received her Ph.D. from Princeton University in Electrical Engineering. From 1994 to 2000, Dr. Mitra was a member of the faculty of the Department of Electrical Engineering at The Ohio State University, Columbus, Ohio. In 2001, she joined the Department of Electrical Engineering at the University of Southern California, Los Angeles, where she is currently a Professor. Dr. Mitra has been an Associate Editor for the following IEEE publications: Transactions on Information Theory (2007-2011), Journal of Oceanic Engineering (2006-2011), and Transactions on Communications (1996-2001). Dr. Mitra served two terms as a member of the IEEE Information Theory Society's Board of Governors (2002-2007).   She is the  recipient of:  USC Center for Excellence in Research Fellowship (2010-2013), Best Applications Paper Award Ð 2009 International Conference on Distributed Computing in Sensor Systems, the Viterbi School of Engineering DeanÕs Faculty Service Award (2009),  USC Mellon Mentoring Award (2008), IEEE Fellow  (2007), Texas Instruments Visiting  Professor (Fall 2002,  Rice University), 2001 Okawa Foundation Award, 2000 Lumley Award  for Research (OSU College of Engineering), 1997 MacQuigg Award for  Teaching (OSU College of Engineering), 1996 National Science  Foundation (NSF) CAREER Award.  She has co-chaired: (technical program)  2012 International Conference on Signal Processing and Communications, Bangalore India, (general) first ACM Workshop on Underwater Networks at Mobicom 2006, Los Angeles, CA and  the (technical) IEEE  Communication Theory Symposium at ICC 2003 in Anchorage, AK.  Dr. Mitra was the tutorials Chair for IEEE ISIT 2007 in Nice, France and the Finance Chair for IEEE ICASSP 2008 in Las Vegas, NV. Dr. Mitra has held visiting appointments at: the Delft University of Technology, Stanford University, Rice University, and the Eurecom Institute.  She served as co-Director of the Communication Sciences Institute at the University of Southern California from 2004-2007.  
\end{biographynophoto}

\begin{biographynophoto}{Michele Zorzi} (F'07) was born in Venice, Italy, in 1966. He received his Laurea degree and Ph.D.\ in electrical engineering from the University of Padova, Italy, in 1990 and 1994, respectively. During academic year 1992/93, he was on leave at the University of California, San Diego (UCSD) attending graduate courses and doing research on multiple access in mobile radio networks. In 1993, he joined the faculty of the Dipartimento di Elettronica e Informazione, Politecnico di Milano, Italy. After spending three years with the Center for Wireless Communications at UCSD, in 1998 he joined the School of Engineering of the University of Ferrara, Italy, and in 2003 joined the Department of Information Engineering of the University of Padova, Italy, where he is currently a Professor. His present research interests include performance evaluation in mobile communications systems, random access in mobile radio networks, ad hoc and sensor networks, energy constrained communications protocols, cognitive networks, and underwater communications and networking.

He was Editor-in-Chief of the \textsc{IEEE Wireless Communications} magazine from 2003 to 2005 and Editor-in-Chief of the \textsc{IEEE Transactions on Communications} from 2008 to 2011, and serves on the Editorial Board of the \textsc{Wiley Journal of Wireless Communications and Mobile Computing}. He was also guest editor for special issues in \textsc{IEEE Personal Communications} and \textsc{IEEE Journal on Selected Areas in Communications}. He served as a Member-at-Large of the Board of Governors of the IEEE Communications Society from 2009 to 2011.
\end{biographynophoto}
\end{document}